\documentclass{tlp}

\usepackage{latexsym}
\usepackage{xspace}
\usepackage{amsmath}
\usepackage{ifthen}
\usepackage{hyperref}

\newcommand{\caso}[2][Case]{\par\medskip\ifthenelse{\equal{}{#1}}{%
    \emph{#2}}{\emph{#1#2}}.}
\newenvironment{case}[2][Case ]{\caso[#1]{#2}}{}

\newcounter{clause}
\def\theclause{c\arabic{clause}}

\newenvironment{clause}{\begin{tabbing}
xxx\=xxx\=xxx\=\+\kill}%
{\end{tabbing}}

\newcommand{\non}{\mathnormal{\sim}}
\newcommand{\TD}{\mathcal{T}_D}
\newcommand{\seq}[2][l]{\ensuremath{#2_{1},\dots,#2_{#1}}\xspace}
\newcommand{\mt}[1]{\mathtt{#1}}
\newcommand{\set}[1]{\{#1\}}

\newcommand{\definitely}{\mathop{\mathtt{definitely}}}
\newcommand{\defeasibly}{\mathop{\mathtt{defeasibly}}}
\newcommand{\strict}{\mathop{\mathtt{strict}}}
\newcommand{\Not}{\mathop{\mathtt{not}}}
\newcommand{\Bf}{\mathbf{f}}
\newcommand{\Bt}{\mathbf{t}}
\newcommand{\plg}{\mathrel{\mbox{\tt:\!-}}}
\newcommand{\fact}{\mathop{\mathtt{fact}}}%
\newcommand{\drule}{\mathop{\mathtt{rule}}}
\newcommand{\suprule}{\mathop{\mathtt{supportive\_rule}}}
\newcommand{\overruled}{\mathop{\mathtt{overruled}}}
\newcommand{\defeated}{\mathop{\mathtt{defeated}}}

\newenvironment{Clause}{\refstepcounter{clause}%
\begin{tabbing}
cxxxx\=xxx\=xxx\=\kill
\theclause\>\+}%
{\end{tabbing}}

\newcommand{\M}{\ensuremath{\mathcal{M}}\xspace}

\newcommand{\pd}[1]{+\partial #1}
\newcommand{\md}[1]{-\partial #1}
\newcommand{\PD}[1]{+\Delta #1}
\newcommand{\MD}[1]{-\Delta #1}

\newcommand{\T}{{\cal T}}

\newtheorem{theorem}{Theorem}[section]
\newtheorem{lemma}{Lemma}[section]

\newtheorem{propn}{Proposition}[section]
\newtheorem{auxexample}{Example}[section]
\newenvironment{example}{\begin{auxexample}\em }{\ $\Box$\end{auxexample}}

\submitted{20 September 2004}
\revised{12 October 2005}
\accepted{8 November 2005}

\begin{document}

\title{Embedding Defeasible Logic into Logic Programming}

\author[G. Antoniou, D. Billington, G. Governatori and M.J. Maher]{
  Grigoris Antoniou \\
\small Institute of Computer Science, FORTH, Greece\\
\small antoniou@ics.forth.gr
\and
David Billington\\
\small School of ICT, Griffith University, Australia \\
\small d.billington@griffith.edu.au
\and
Guido Governatori\\
\small School of ITEE, University of Queensland, Australia\\
\small guido@itee.uq.edu.au
\and
Michael J. Maher\\
\small National ICT Australia, 
  c/o UNSW, Australia\\
\small mmaher@cse.unsw.edu.au}

\date{}

\maketitle

\begin{abstract}
  Defeasible reasoning is a simple but efficient approach to
  nonmonotonic reasoning that has recently attracted considerable
  interest and that has found various applications. Defeasible logic
  and its variants are an important family of defeasible reasoning
  methods. So far no relationship has been established between
  defeasible logic and mainstream nonmonotonic reasoning approaches.
  
  In this paper we establish close links to known semantics of logic
  programs. In particular, we give a translation of a defeasible
  theory $D$ into a meta-program $P(D)$. We show that under a
  condition of decisiveness, the defeasible consequences of $D$
  correspond exactly to the sceptical conclusions of $P(D)$ under the
  stable model semantics. Without decisiveness, the result holds only
  in one direction (all defeasible consequences of $D$ are included in
  all stable models of $P(D)$). If we wish a complete embedding for
  the general case, we need to use the Kunen semantics of $P(D)$,
  instead.
\end{abstract}


\section{Introduction}

Defeasible reasoning is a nonmonotonic reasoning \cite{MT93}
approach in which the gaps due to incomplete information are
closed through the use of defeasible rules that are usually
appropriate. Defeasible logics were introduced and developed by
Nute \citeyear{Nute94} over several years.  These logics perform
defeasible reasoning, where a conclusion supported by a rule might
be overturned by the effect of another rule. Roughly, a
proposition $p$ can be defeasibly proved ($+\partial p$) only when
a rule supports it, and it has been demonstrated that no
applicable rule supports $\neg p$; this demonstration makes use of
statements $-\partial q$ which mean intuitively that an attempt to
prove $q$ defeasibly has failed finitely. These logics also have a
monotonic reasoning component, and a priority on rules. One
advantage of Nute's design was that it was aimed at supporting
efficient reasoning, and in our work we follow that philosophy.

Defeasible reasoning has recently attracted considerable interest.
Its use in various application domains has been advocated, including
the modelling of regulations and business rules
\cite{leora,Antoniou99}, modelling of contracts
\cite{Grosof99,Grosof:ECRA,Guido05}, legal reasoning \cite{Prakken,icail05},
agent negotiations \cite{icail01}, modelling of agents and agent
societies \cite{GovRot:deon04,ai03v,adc04grs}, and applications to the
Semantic Web \cite{BasAntVlah:04:DrDevice,Ant:ICSW:02:Nonmonotonic}.
In fact, defeasible reasoning (in the form of courteous logic programs
\cite{Grosof97,Grosof99}) provides a foundation for IBM's Business
Rules Markup Language and for current W3C activities on rules
\cite{Grosof:03:Sweetdeal,Grosof:02:Sweetjess}. In addition,
defeasible theories, describing policies of business activities, can
be mined efficiently from appropriate datasets \cite{adc03}.
Therefore defeasible reasoning is a promising subarea in nonmonotonic
reasoning as far as applications and integration to mainstream IT is
concerned.

Recent theoretical work on defeasible logics has: (i) established some
relationships to other logic programming approaches without negation
as failure \cite{Antoniou00a}; (ii) analysed the formal properties of
these logics \cite{Antoniou00c,Maher00,Maher01} as well as formal
semantics for them in form of model theoretic semantics \cite{Maher02}
and argumentation semantics \cite{argumentation}, and (iii) has
delivered efficient implementations \cite{Tools}.

However the problem remains that defeasible logic is not firmly
linked to the mainstream of nonmonotonic reasoning, in particular
the semantics of logic programs. This paper aims at resolving this
problem. We use the  translation of a defeasible theory $D$ into a
logic meta-program ${\cal M}$ proposed in \cite{Maher99}. For this
translation we can show that, for the propositional case:
\begin{quote}
  $p$ is defeasibly provable in $D$ $\Longleftrightarrow$ $p$ is
  included in all stable models of ${\cal M}$. \hfill $(*)$
\end{quote}
However this result can only be shown under the additional
condition of {\em decisiveness}: the absence of cycles in the atom
dependency graph.

If we wish to drop decisiveness, $(*)$ holds only in one direction,
from left to right. We show that if we wish the equivalence in the
general case, we need to use another semantics for logic programs,
namely Kunen semantics \cite{Kunen}.  
In addition the possibility of cycles in the atom dependency graph of
a defeasible theory prevents Defeasible Logic to be characterised by
well-founded semantics \cite{vanGelder}. It is possible to modify
Defeasible Logic to accommodate well-founded semantics \cite{Maher99}
even if this results in a more expensive computational model.

The paper is organised as follows. Sections 2 and 3 present the basics
of defeasible logic and logic programming semantics, respectively.
Section 4 presents our translation of defeasible theories in logic
programs, while section 5 contains the main results.

\section{Defeasible Logic}

\subsection{A Language for Defeasible Reasoning}

A defeasible theory (a knowledge base in defeasible logic)
consists of five different kinds of knowledge: facts, strict
rules, defeasible rules, defeaters, and a superiority relation.

{\em Facts} are literals that are treated as known knowledge
(given or observed facts of a case).

\smallskip\noindent
{\em Strict rules} are rules in the classical sense: whenever the
premises are indisputable (e.g.\ facts) then so is the conclusion.
An example of a strict rule is ``Emus are birds''. Written
formally:
\[
emu(X) \rightarrow bird(X).
\]

\smallskip\noindent
{\em Defeasible rules} are rules that can be defeated by contrary
evidence. An example of such a rule is ``Birds typically fly'';
written formally:
\[
bird(X) \Rightarrow flies(X).
\]
The idea is that if we know that something is a bird, then we may
conclude that it flies, {\em unless there is other, not inferior,
evidence suggesting that it may not fly}.

\smallskip\noindent
{\em Defeaters} are rules that cannot be used to draw any
conclusions. Their only use is to prevent some conclusions. In
other words, they are used to defeat some defeasible rules by
producing evidence to the contrary. An example is ``If an animal
is heavy then it might not be able to fly''. Formally:
\[
heavy(X) \leadsto \neg flies(X).
\]
The main point is that the
information that an animal is heavy is not sufficient evidence to
conclude that it doesn't fly. It is only evidence against the
conclusion that a heavy animal flies. In other words, we don't
wish to conclude $\neg flies$ if $heavy$, we simply want to
prevent a conclusion $flies$.

\smallskip\noindent
The {\em superiority relation} among rules is used to define
priorities among rules, that is, where one rule may override the
conclusion of another rule.  For example, given the defeasible
rules
\[
\begin{array}{lrl}
r: & bird(X) & \Rightarrow flies(X) \\
r':  & brokenWing(X) & \Rightarrow \neg flies(X) \\
\end{array}
\]
which contradict one another, no conclusive decision can be made about
whether a bird with broken wings can fly. But if we introduce a
superiority relation $>$ with $r'>r$, with the intended meaning that
$r'$ is strictly stronger than $r$, then we can indeed conclude that
the bird cannot fly.

It is worth noting that, in defeasible logic, priorities are {\em
local} in the following sense: Two rules are considered to be
competing with one another only if they have complementary heads.
Thus, since the superiority relation is used to resolve conflicts
among competing rules, it is only used to compare rules with
complementary heads; the information $r>r'$ for rules $r,r'$
without complementary heads may be part of the superiority
relation, but has no effect on the proof theory.

\subsection{Formal Definition}

In this paper we restrict attention to essentially propositional
defeasible logic. Rules with free variables are interpreted as
rule schemas, that is, as the set of all ground instances; in such
cases we assume that the Herbrand universe is finite.  We assume
that the reader is familiar with the notation and basic notions of
propositional logic. If $q$ is a literal, $\non q$ denotes the
complementary literal (if $q$ is a positive literal $p$ then $\non
q$ is $\neg p$; and if $q$ is $\neg p$, then $\non q$ is $p$).

Rules are defined over a {\em language} (or {\em signature})
$\Sigma$, the set of propositions (atoms) and labels that may be
used in the rule.

A {\em rule} $r:A(r)\hookrightarrow C(r)$ consists of its unique
\emph{label} $r$, its {\em antecedent} $A(r)$ ($A(r)$ may be
omitted if it is the empty set) which is a finite set of literals,
an arrow $\hookrightarrow$ (which is a placeholder for concrete
arrows to be introduced in a moment), and its {\em head} (or {\em
consequent}) $C(r)$ which is a literal. In writing rules we omit
set notation for antecedents and sometimes we omit the label when
it is not relevant for the context.  There are three kinds of
rules, each represented by a different arrow.  Strict rules use
$\rightarrow$, defeasible rules use $\Rightarrow$, and defeaters
use $\leadsto$.

Given a set $R$ of rules, we denote the set of all strict rules in
$R$ by $R_s$, and the set of strict and defeasible rules in $R$ by
$R_{sd}$. $R[q]$ denotes the set of rules in $R$ with consequent
$q$.

A {\em superiority relation on $R$} is a relation $>$ on $R$.
When $r_1>r_2$, then $r_1$ is called {\em superior} to $r_2$, and
$r_2$ {\em inferior} to $r_1$.  Intuitively, $r_1>r_2$ expresses
that $r_1$ overrules $r_2$, should both rules be applicable. $>$
must be acyclic (that is, its transitive closure must be
irreflexive).

A {\em defeasible theory} $D$ is a triple $(F,R,>)$ where $F$ is a
finite set of facts, $R$ a finite set of rules, and $>$ an acyclic
superiority relation on $R$.

\subsection{Proof Theory}

A {\em conclusion} of a defeasible theory $D$ is a tagged literal.
A conclusion has one of the following four forms:

\begin{itemize}
\item $+\Delta q$, which is intended to mean that the literal $q$ is
  definitely provable, using only strict rules.
  
\item $-\Delta q$, which is intended to mean that $q$ is provably not
  definitely provable (finite failure).
  
\item $+\partial q$, which is intended to mean that $q$ is defeasibly
  provable in $D$.
\item $-\partial q$ which is intended to mean that we have proved that
  $q$ is not defeasibly provable in $D$.
\end{itemize}

Provability is defined below. It is based on the concept of a {\em
derivation} (or {\em proof})  in $D=(F,R,>)$. A derivation is a
finite sequence $P=P(1),\ldots, P(n)$ of tagged literals
satisfying the following conditions. The conditions are
essentially inference rules phrased as conditions on proofs.
$P(1..i)$ denotes the initial part of the sequence $P$ of length
$i$.

\begin{tabbing}
\=7890\=1234\=5678\=9012\=3456\=\kill

\>$+\Delta$: If $P(i+1) = +\Delta q$ then either \\
\>\>\> $q\in F$ or \\
\>\>\> $\exists r\in R_s[q] \ \forall a\in A(r): +\Delta a \in
P(1..i)$
\end{tabbing}

That means, to prove $+\Delta q$ we need to establish a
proof for $q$ using facts and strict rules only. This is a
deduction in the classical sense -- no proofs for the negation of
$q$ need to be considered (in contrast to defeasible provability
below, where opposing chains of reasoning must be taken into
account, too). 

\begin{tabbing}
\=7890\=1234\=5678\=9012\=3456\=\kill

\> $-\Delta$: If $P(i+1)=-\Delta q$ then \\
\>\>\> $q\not\in F$ and \\
\>\>\> $\forall r\in R_{s}[q] \ \exists a\in A(r): -\Delta a \in
P(1..i)$
\end{tabbing}

To prove $-\Delta q$, i.e.\ that $q$ is not definitely
provable, $q$ must not be a fact. In addition, we need to
establish that every strict rule with head $q$ is {\em known to
be} inapplicable. Thus for every such rule $r$ there must be at
least one antecedent $a$  for which we have established that $a$
is not definitely provable ($-\Delta a$).

\begin{tabbing}
\=7890\=1234\=5678\=9012\=3456\=\kill
\>$+\partial$: \> If $P(i+1)=+\partial q$ then either \\
\>\>(1) $+\Delta q \in P(1..i)$ or \\
\>\>(2) \> (2.1) $\exists r\in R_{sd}[q]\ \forall a \in
A(r): +\partial a\in P(1..i)$ and \\
\>\>\>(2.2) $-\Delta\non q \in P(1..i)$ and \\
\>\>\>(2.3) $\forall s \in R[\non q]$ either \\
\>\>\>\>(2.3.1) $\exists a\in A(s): -\partial a\in P(1..i)$ or \\
\>\>\>\>(2.3.2) $\exists t\in R_{sd}[q]$ such that \\
\>\>\>\>\> $\forall a\in A(t): +\partial a\in P(1..i)$ and $t>s$
\end{tabbing}

Let us illustrate this definition. To show that $q$ is provable
defeasibly we have two choices: (1) We show that $q$ is already
definitely provable; or (2) we need to argue using the defeasible
part of $D$ as well. In particular, we require that there must be
a strict or defeasible rule with head $q$ which can be applied
(2.1). But now we need to consider possible ``attacks'',
that is, reasoning chains in support of $\non q$. To be more
specific: to prove $q$ defeasibly we must show that $\non q$ is
not definitely provable (2.2). Also (2.3) we must consider the set
of all  rules which are not known to be inapplicable and which
have head $\non q$ (note that here we consider defeaters, too,
whereas they could not be used to support the conclusion $q$; this
is in line with the motivation of defeaters given above).
Essentially each such rule $s$ attacks the conclusion $q$. For $q$
to be provable, each such rule $s$ must be counterattacked by a
rule $t$ with head $q$ with the following properties: (i) $t$ must
be applicable at this point, and (ii) $t$ must be stronger than
(i.e. superior to) $s$. Thus each attack on the conclusion $q$
must be counterattacked by a stronger rule.

\begin{tabbing}
\=7890\=1234\=5678\=9012\=3456\=\kill

\>$-\partial$: \> If $P(i+1)=-\partial q$ then  \\
\>\>(1) $-\Delta q \in P(1..i)$ and \\
\>\>(2) \> (2.1) $\forall r\in R_{sd}[q] \ \exists a \in
A(r): -\partial a\in P(1..i)$ or \\
\>\>\>(2.2) $+\Delta\non q \in P(1..i)$ or \\
\>\>\>(2.3) $\exists s \in R[\non q]$ such that \\
\>\>\>\>(2.3.1) $\forall a\in A(s): +\partial a\in P(1..i)$ and \\
\>\>\>\>(2.3.2) $\forall t\in R_{sd}[q]$ either \\
\>\>\>\>\> $\exists a\in A(t): -\partial a\in P(1..i)$ or $t\not>
s$
\end{tabbing}

To prove that $q$ is not defeasibly provable, we must
first establish that it is not definitely provable. Then we must
establish that it cannot be proven using the defeasible part of
the theory. There are three possibilities to achieve this: either
we have established that none of the (strict and defeasible) rules
with head $q$ can be applied (2.1); or $\non q$ is definitely
provable (2.2); or there must be an applicable rule $s$ with head
$\non q$ such that no applicable rule $t$ with head $q$ is
superior to $s$. 

In general the inference conditions for a negative proof tag (i.e.,
$-\Delta$, $-\partial$) explore all the possibilities to derive a
literal (with a given proof strength) before stating that the literal
is not provable (with the same proof strength). Thus conclusions with
these tags are the outcome of a constructive proof that the
corresponding positive conclusion cannot be obtained.  As a result,
there is a close relationship between the inference rules for
$+\partial$ and $-\partial$, (and also between those for $+\Delta$ and
$-\Delta$). The structure of the inference rules is the same, but the
conditions are negated in some sense. To be more precise the inference
conditions for a negative proof tag are derived from the inference
conditions for the corresponding positive proof tag by applying the
Principle of Strong Negation introduced in \cite{Antoniou00b}.  The
strong negation of a formula is closely related to the function that
simplifies a formula by moving all negations to an innermost position
in the resulting formula and replaces the positive tags with the
respective negative tags and vice-versa.

The elements of a derivation are called {\em lines} of the
derivation. We say that a tagged literal $L$ is {\em provable} in
$D=(F,R,>)$, denoted by $D\vdash L$, iff there is a derivation $P$ in
$D$ such that $L$ is a line of $P$.

Defeasible logic is closely related to several non-monotonic
logics \cite{Antoniou99}. In particular, the ``directly
skeptical'' semantics of non-monotonic inheritance networks
\cite{Horty87} can be considered an instance of inference in $DL$
once an appropriate superiority relation, derived from the
topology of the network, is fixed \cite{Billington90}.

A defeasible theory $D$ is {\em coherent}\footnote{Notice that here
  coherent has a different meaning than other works on logic
  programming for example \cite{Alferes1,Alferes2}.} if for no literal
$p$ both $D\vdash+\Delta p$ and $D\vdash-\Delta p$, and
$D\vdash+\partial p$ and $D\vdash-\partial p$; and {\em relatively
  consistent} if whenever $D\vdash +\partial p$ and $D\vdash
+\partial\non p$, for some $p$, then also $D\vdash +\Delta p$ and
$D\vdash+\Delta\non p$.  Intuitively, coherence says that no literal
is simultaneously provable and unprovable.  Consistency says that a
literal and its negation can both be defeasibly provable only when it
and its negation are definitely provable; hence defeasible inference
does not introduce inconsistency.  A logic is coherent (relatively
consistent) if the meaning of each theory of the logic, when expressed
as an extension (i.e., when we consider the set of all the
consequences of the theory), is coherent (relatively consistent), that
is it is not possible to prove a formula and its negation unless the
monotonic part of the theory proves them.

\begin{propn}[\citeNP{Billington93}]
  Defeasible logic is coherent and relatively consistent.
\end{propn}

Consistency and coherence address the issue whether and how it is
possible to derive ``conflicts'' from defeasible theories. On the
other side we can ask under which conditions defeasible theories are
\emph{complete}, in the sense that for every literal in the theory it is
possible to decide whether the literal is provable/non provable
from the theory. 

In the rest of this section we will study conditions under which it is
possible to guarantee completeness of a defeasible theory.

Given a defeasible theory $D=(F,R,>)$ a literal $q$ is \emph{strictly
  unknowable in $D$} iff $D\not\vdash+\Delta q$ and
$D\not\vdash-\Delta q$. A literal is \emph{defeasibly unknowable in
  $D$} iff $D\not\vdash+\partial q$ and $D\not\vdash-\partial
q$. A literal is \emph{unknowable} in $D$ iff it is either strictly
unknowable in $D$ or defeasibly unknowable in $D$. 

The \emph{dependency graph} of $D$, $DG(D)$, is the directed graph
defined as follows: the set of points of $DG(D)$ is $\set{\set{q,\non
    q}: q \text{ is a literal in }D}$. The set of arcs of $DG(D)$ is
$\set{(\set{b,\non b}, \set{a,\non a})| \exists r\in R_s[b]:
  a\in A(r)}$. Let $U(D)$ be the subgraph of $DG(D)$ restricted to
the literals that are unknowable in $D$, i.e., the set of points of
$U(D)$ is $\set{\set{q,\non q}: q \text{ is unknowable in }D}$. 

The next Lemmata show the mutual relationships among unknowable
literals in a Defeasible theory, and general properties of graphs of
unknowable literals.
\begin{lemma}\label{lem:degree}
  Let $D=(F,R,>)$ be a Defeasible Theory. %Then
  The out-degree of $U(D)$ is at least $1$.
\end{lemma}
\begin{proof}
  We have to consider two cases: strictly unknowable literals and
  defeasibly unknowable literals.

  Let $q$ be a strictly unknowable literal, then there is a point
  $\set{q,\non q}$ in $U(D)$. Then, by definition, $D\not\vdash+\Delta
  q$ and $D\not\vdash-\Delta q$.\footnote{The case of $\non q$ is
    identical.} Hence $q\notin F$ and $\forall r\in R_s[q]\exists a\in
  A(r)$ such that $D\not\vdash+\Delta a$. Since $D\not\vdash-\Delta q$
  and $q\notin F$, then $\exists r\in R_s[q]$ such that $\forall a\in
  A(r)$ $D\not\vdash -\Delta a$. So $\exists r\in R_s[q]$ $\exists
  a\in A(r)$ such that $D\not\vdash +\Delta a$ and $D\not\vdash-\Delta
  a$. Thus $(\set{q,\non q}, \set{a,\non a})$ is an arc in $U(D)$.

  If $q$ is a defeasibly unknowable literal then we reason as
  follows: let $\set{q,\non q}$ be the point in $U(D)$ corresponding to
  $q$. Then $D\not\vdash +\partial q$ and $D\not\vdash-\partial q$.

  Since $D\not\vdash +\partial q$ we have the following:
  \begin{tabbing}
    a1) $D\not\vdash +\Delta q$ and\\
    a2) \=one of the following three holds:\+\\
    a2.1) $\forall r\in R_{sd}[p]\exists a\in A(r): D\not\vdash +\partial
        a$; or\\
    a2.2) $D\not\vdash -\Delta \non q$; or\\
    a2.3) \=$\exists s\in R[\non q]$ such that \+\\
      a2.3.1) $\forall a\in A(s)\ D\not\vdash -\partial a$, and\\
      a2.3.2) \=$\forall t\in R_{sd}[q]$ either\+\\
            $\exists a\in A(t): D\not\vdash +\partial a$ or $t\not>s$.
  \end{tabbing}
  Since $D\not\vdash-\partial q$ we have the following.
  \begin{tabbing}
  b1) $D\not\vdash-\Delta q$; or\\
  b2) \=All of the following three hold:\+\\
    b2.1) $\exists r\in R_{sd}[q]\forall a\in A(r)\ D\not\vdash
          -\partial a$, and\\
    b2.2) $D\not\vdash +\Delta \non q$, and\\
    b2.3) \=$\exists s\in R[\non q]$ either\+\\
     b2.3.1) $\exists a\in A(s), D\not\vdash +\partial a$; or\\
     b2.3.2) \=$\exists t\in R_{sd}[q]$ such that \+\\
       $\forall a\in A(t), D\not\vdash -\partial a$ and $t>s$.
  \end{tabbing}
  By a1, $D\not\vdash +\Delta q$. If b1 holds then $q$ is strictly
  unknowable and we can repeat the first part of the proof. So
  suppose that b2 holds. 

  If a2.1 holds then by b2.1 $\exists r\in R_{sd}[p]\exists a\in A(r)$
  such that $a$ is defeasibly unknowable in $D$. Thus $(\set{q,\non
  q}, \set{a, \non a})$ is an arc in $U(D)$. If a2.2 holds then by
  b2.2 $\non p$ is a strictly unknowable literal in $D$, and we have
  already proved the property in this case.

  So suppose that a2.3 holds. If b2.3 holds then $\exists s\in R[\non
  p]\exists a\in A(s)$ such that $a$ is a defeasibly unknowable
  literal in $D$. Thus  $(\set{q,\non q}, \set{a, \non a})$ is an
  arc in $U(D)$. So suppose that b2.3.2 holds. Then $t>s$ and so
  a2.3.2 holds. Hence $\exists t\in R_{sd}[q]\exists a\in A(t)$ such
  that $a$ is defeasibly unknowable in $D$. Thus $(\set{q,\non
  q}, \set{a, \non a})$ is an arc in $U(D)$.

  Therefore in all cases the out-degree of $\set{q,\non q}$ is at
  least 1.
\end{proof}
Given a graph, a \emph{walk} is an alternating sequence of vertices
and edges, with each edge being incident to the vertices immediately
preceeding and succeeding it in the sequence.

The set of points in a walk $W$ in the dependency graph of a theory $D$
is denoted by $Points(W)$. A walk $W$ \emph{ends in a cycle} iff $W$ is
finite and the last point of $W$ occurs at least twice. A walk is
\emph{complete} iff either
\begin{enumerate}
\item $Points(W)$ is infinite; or
\item a point in $W$ has out-degree zero; or
\item $W$ ends in a cycle.
\end{enumerate}
\begin{lemma}\label{lem:walks}
  Let $D=(F,R,>)$ be a defeasibly theory. Then the following are equivalent.
  \begin{enumerate}
  \item \label{item:first} There is a literal which is unknowable in $D$.
  \item \label{item:second} $U(D)$ is not empty.
  \item \label{item:third} There is a walk in $U(D)$.
  \item \label{item:fourth} There is a complete walk in $U(D)$.
  \item \label{item:fifth} There is a complete walk in $U(D)$ and
    whenever $W$ is a complete walk in $U(D)$ then either $Points(W)$
    is infinite or $W$ ends in a cycle.
  \end{enumerate}
\end{lemma}
\begin{proof}
  \ref{item:first} and \ref{item:second} are clearly equivalent. If
  \ref{item:second} holds then by Lemma \ref{lem:degree},
  \ref{item:third} holds. If \ref{item:third} holds, then
  \ref{item:fourth} holds since every walk can be extended to a
  complete walk.   By Lemma \ref{lem:degree}, every point in $U(D)$
  has out-degree of at least 1, and so if \ref{item:fourth} holds,
  then \ref{item:fifth} holds. If \ref{item:fifth} holds since walks
  are not empty \ref{item:second} holds.
\end{proof}

A defeasible theory $D$ is called {\em decisive} iff the dependency graph
of $D$ is acyclic.

The following proposition provides a sufficient
condition to determine completeness of a defeasible theory.
 
\begin{theorem} 
If $D$ is decisive, then for each literal $p$:
\begin{itemize}
\item[(a)] either $D\vdash +\Delta p$ or $D\vdash -\Delta p$
\item[(b)] either $D\vdash +\partial p$ or $D\vdash -\partial p$.
\end{itemize}
\end{theorem}
\begin{proof}
  We prove the contrapositive, i.e., suppose there are unknowable literals.
  If there are unknowable literals then by Lemma \ref{lem:walks},
  there is a walk in $U(D)$ such that $Points(W)$ is infinite or $W$
  ends in a cycle. Since there are only finitely many rules in $D$,
  $Points(W)$ is finite, thus $W$ ends in a cycle. Thus $U(D)$ has a
  cycle and $U(D)$ is a subgraph of $DG(D)$, thus $D$ is not decisive. 
\end{proof}

Not every defeasible theory satisfies this property. For example,
in the theory consisting of the single rule
\[
r_1:p\Rightarrow p
\]
neither $-\partial p$ nor $+\partial p$ is provable. 

Notice, however, that there are complete non decisive theories. If we
extend the above theory with the rule
\[
r_2: {} \Rightarrow \neg p
\]
and the superiority relation $>$ is defined as $r_2>r_1$
then we can prove both $-\partial p$ and $+\partial \neg p$, thus the
resulting theory is not decisive, but complete.

\subsection{A Bottom-Up Characterization of Defeasible Logic}

The proof theory provides the basis for a top-down (backward-chaining)
implementation of the logic. In fact the {\em Deimos} system
\cite{Tools} is based directly on the proof theory described above.
However, there are advantages to a bottom-up (forward-chaining)
implementation.  In particular this presentation of Defeasible Logic
provides a both a set theoretic and a declarative computational model
of the logic compared to the procedural characterisation of the
top-down definitions. This allows us to describe a defeasible theory
as an extension (i.e., set of all conclusions provable from it) and to
deal with finite as well as infinite theories. This is not possible in
the other approach since derivations are required to be finite
sequences of (tagged) literals and they are described in term of
combinatorial constructions.  Furthermore, a bottom-up definition of
the logic provides a bridge to later considerations.  For these
reasons we now provide a bottom-up definition of Defeasible Logic.

We associate with $D$ an operator $\T_D$ which works on 4-tuples of
sets of literals.  We call such 4-tuples an {\em extension}.

\begin{tabbing}
123\=123456\=123456\=123456\=123456\=123456\=\kill

\>$\T_D(\PD{}, \MD{}, \pd{}, \md{}) =
(\PD{}', \MD{}', \pd{}', \md{}')$ where \\
\\
$\PD{}' = F  ~ \cup ~ \{q ~|~
\exists r\in R_s[q] \ A(r) \subseteq \PD{}
\} $ \\
\\
$\MD{}' = \MD{}  ~ \cup ~ (\{q ~|~
\forall r\in R_s[q] \ A(r) \cap \MD{} \neq \emptyset
\} ~ - ~ F) $ \\

\> \\

$\pd{}' = \PD{} ~ \cup ~$
$\{q ~|~
\exists r\in R_{sd}[q] \ A(r) \subseteq \pd{},$ \\
\>\>\> $\non q \in \MD{},$ and \\
\>\>\> $\forall s \in R[\non q]$ either \\
\>\>\>\>$A(s) \cap \md{} \neq \emptyset$, or \\
\>\>\>\> $\exists t\in R[q]$ such that \\
\>\>\>\>\> $A(t) \subseteq \pd{}$ and $t>s \}$ \\

\> \\

 $\md{}' = \{q \in \MD{} ~|~
\forall r\in R_{sd}[q] \ A(r) \cap \md{} \neq \emptyset$, or \\
\>\>\> $\non q \in \PD{}$, or \\
\>\>\> $\exists s \in R[\non q]$ such that
$A(s) \subseteq \pd{}$ and \\
\>\>\> $\forall t\in R[q]$ either \\
\>\>\>\> $A(t) \cap \md{} \neq \emptyset$, or \\
\>\>\>\> $ t\not> s \} $

\end{tabbing}

The set of extensions forms a complete lattice under the pointwise
containment ordering\footnote{$(a_1, a_2, a_3, a_4) \leq (b_1, b_2,
  b_3, b_4)$ iff $a_i \subseteq b_i$ for $i=1,2,3,4$.}, with $\bot =
(\emptyset, \emptyset, \emptyset, \emptyset)$ as its least element.
The least upper bound operation is the pointwise union\footnote{Given
  two $n$-tuple of sets $a=(a_1,\dots, a_n)$ and $b=(b_1,\dots,b_2)$
  the pointwise union of $a$ and $b$ is defined as follows: $a\cup b=
  (a_1\cup b_1, \dots, a_n\cup b_n)$.}, which is
represented by $\cup$.

The sequence of repeated applications of $T_D$ to $\bot$, called
the {\em Kleene sequence} of $T_D$, is defined as follows:
\begin{itemize}
\item $T_D \uparrow 0 = \bot$;
\item $T_D \uparrow (\alpha + 1) = T_D( T_D \uparrow \alpha )$;
\item $T_D \uparrow \alpha = \bigcup_{\beta < \alpha} T_D \uparrow \beta$
if $\alpha$ is a limit ordinal.
\end{itemize}

\begin{propn} \label{fixpoint}
  $\T_D$ is monotonic and the Kleene sequence from $\bot$ is
  increasing.  Thus the limit $F = (\PD{}_F, \MD{}_F, \pd{}_F,
  \md{}_F)$ of all finite elements in the sequence exists, and $\T_D$
  has a least fixpoint $L = (\PD{}_L, \MD{}_L, \pd{}_L, \md{}_L)$.
  When $D$ is a finite propositional defeasible theory $F = L$.
\end{propn}
\begin{proof}
We prove by induction that $\mathcal{T}_D$ is pointwise monotonic. The
other properties follow from standard and well-know set theoretic
arguments. 

The inductive base is trivial since the elements of $\bot$ are
$\emptyset$. 

For the inductive step we have four cases, where the inductive
hypothesis amounts to: $+\Delta^{n-1}\subseteq +\Delta^n$,
$-\Delta^{n-1}\subseteq -\Delta^n$, $+\partial^{n-1}\subseteq +\partial^n$
and $-\partial^{n-1}\subseteq -\partial^n$.

\begin{case}{$+\Delta$}
  Let us investigate the reasons why $p\in+\Delta^{n}$: if $p\in F$,
  then, trivially, for all $m$ $p\in +\Delta^m$; hence $p\in
  +\Delta^{n+1}$. Otherwise $\exists r\in R_s[p]$ such that
  $A(r)\subseteq +\Delta^{n-1}$. By inductive hypothesis
  $A(r)\subseteq +\Delta^n$. Therefore $p\in+\Delta^{n+1}$.
\end{case}

\begin{case}{$-\Delta$}
  Trivial since $-\Delta^{n+1}=-\Delta^n\cup S$, for some set of
  literals $S$.
\end{case}

\begin{case}{$+\partial$}
  If $p\in+\partial^n$ because $p\in+\Delta^{n-1}$, then by inductive
  hypothesis $p\in+\Delta^n$ and thus $p\in+\partial^{n+1}$. Otherwise
  $p\in+\partial^n$ if (i) $\exists r\in R_{sd}[p]$ such that
  $A(r)\subseteq +\partial^{n-1}$ and (ii) $\non p\in -\Delta^{n-1}$ and
  $\forall s\in R[\non p]$ either (iii) $A(s)\cap -\partial^{n-1}\neq
  \emptyset$ or (iv) $\exists t\in R[p]$ such that $A(t)\subseteq
  +\partial^{n-1}$ and $t>s$. By inductive hypothesis, if (i) then
  $A(r)\subseteq +\partial^n$, if (ii) then $\non p\in -\Delta^n$, if
  (iii) then $A(s)\cap -\partial^n\neq\emptyset$, and if (iv) then
  $A(t)\subseteq +\partial^n$. Therefore every time the conditions for
  $p$ being in $+\partial^n$ are satisfied, so are those for $p$
  being in $+\partial^{n+1}$.
\end{case}

\begin{case}{$-\partial$}
  First of all, by definition, $-\partial^n\subseteq -\Delta^n$, then
  we have three possibilities to add a literal $p$ to
  $-\partial^{n+1}$. (i) $\forall r\in R_{sd}[p] A(r)\cap
  -\partial^{n-1}\neq \emptyset$. by inductive hypothesis
  $-\partial^{n-1}\subseteq -\partial^n$, thus $A(r)\cap
  -\partial^n\neq \emptyset$, hence $p\in -\partial^{n+1}$. (ii) $\non
  p\in +\Delta^{n-1}$, but by inductive hypothesis
  $+\Delta^{n-1}\subseteq +\Delta^n$, thus $p\in -\partial^{n+1}$.
  (iii) $\exists s\in R[\non p]$ such that $A(s)\subseteq
  +\partial^{n-1}$ and $\forall t\in R[p]$ either $A(t)\cap
  -\partial^{n-1}\neq \emptyset$ or $t\not> s$. By inductive
  hypothesis $+\partial^{n-1}\subseteq +\partial^n$ and
  $-\partial^{n-1}\subseteq -\partial^{n}$. Therefore, also in this
  case, $p\in -\partial^{n+1}$. Hence $p\in -\partial^{n+1}$.
\end{case}
\end{proof}

The extension $F$ captures exactly the inferences described
in the proof theory.

\begin{theorem} \label{butd} Let $D$ be a finite propositional
  defeasible theory, $q$ a literal and
  \[
  F=(\PD{}_F, \MD{}_F,\pd{}_F,\md{}_F)
  \]
  is the limit of all finite elements of the Kleene
  sequence from $\bot$ via $\T_D$.

Then:
\begin{itemize}
\item $D\vdash \PD{q}$ iff $q\in\PD{}_F$
\item $D\vdash \MD{q}$ iff $q\in\MD{}_F$
\item $D\vdash \pd{q}$ iff $q\in\pd{}_F$
\item $D\vdash \md{q}$ iff $q\in\md{}_F$
\end{itemize}
\end{theorem}

\begin{proof}
  We prove the theorem by induction on the length of derivations in
  one direction and on the number of iterations of the operator $\TD$
  in the other.

%%%%%%%%%%%%%%%%%%%%%%%%%%%%%%%%%%%%%%%%%%%%%%%%%%
% proof of +\Delta
%%%%%%%%%%%%%%%%%%%%%%%%%%%%%%%%%%%%%%%%
\begin{case}[\textbf{Case} $+\Delta$, ]{Inductive base $\Rightarrow$}
$P(1)=+\Delta q$. This means that either $q\in F$ or that there exists
a rule $r\in R_s[q]$ such that $A(r)=\emptyset$. In both cases $q\in
+\Delta^1_F$. In the first case by definition, in the second since
$\emptyset\subseteq +\Delta^0=\emptyset$. By the monotonicity of
$\TD$, $q\in +\Delta_F$.
\end{case}

\begin{case}[Inductive step]{}
We assume that the theorem holds for proofs of length up to $n$, and
$P(n+1)=+\Delta q$. Here we consider only the cases different from the
inductive base. Thus there exists a rule $r\in R_s[q]$ such that
$\forall a\in A(r)$, $+\Delta a\in P(1..n)$. By inductive hypothesis
$a\in +\Delta_F$. Let $+\Delta^m$ be the minimal set of literals in
the Kleene sequence defined by $\TD$ containing all such $a$s. Clearly
$A(r)\subseteq \Delta^{m+1}$. Hence, by the monotonicity of $\TD$,
$q\in+\Delta_F$. 
\end{case}

\begin{case}[Inductive base $\Leftarrow$]{}
If $q\in +\Delta^1$, then either $q\in F$ or $\exists r\in R_s[q]:
A(r)\subseteq +\Delta^0$, that is $A(r)=\emptyset$. In both case we
have that $+\Delta q$ is a single line proof of $q$.  
\end{case}

\begin{case}[Inductive step]{}
We have that $q\in +\Delta^{n+1}$ and the property holds up to
$+\Delta^n$. If $q\in+\Delta^{n+1}$ because $q\in F$, then, as in the
previous case, $+\Delta q$ is a single line proof of $q$. Otherwise
$\exists r\in R_s[q]:A(r)\subseteq +\Delta^n$. By inductive hypothesis
$\forall a_i\in A(r)$, $D\vdash +\Delta a_i$. Let $\seq{a}$ be an
enumeration of the literals in $A(r)$, and let $P(a_i)$ be a proof of
$a_i$. We concatenate the $P(a_i)$s and we append $+\Delta q$ at the
end. It is immediate to verify that the sequence thus obtained is a
proof of $+\Delta q$.
\end{case}

%%%%%%%%%%%%%%%%%%%%%%%%%%%%%%%%%%%%%%%%%%%%%%%%%%
% proof of -\Delta
%%%%%%%%%%%%%%%%%%%%%%%%%%%%%%%%%%%%%%%%%%%%%%%%%%
\begin{case}[\textbf{Case} $-\Delta$, ]{Inductive base $\Rightarrow$}
$P(1)=-\Delta q$ iff $p\notin F$ and $\neg \exists r\in R_s$ such that
$C(r)=q$. On the other hand $-\Delta^0=\emptyset$, so the set of
literals satisfying $\forall s\in R_s[q]: A(s)\cap
-\Delta^0\neq\emptyset$ is the set of literals not appearing as the
consequent in a strict rule in $D$. Moreover the definition of 
relative complement gives us 
\[
-\Delta^1=\{p| \neg\exists r\in R_s \wedge p\notin F\}
\]
Therefore $q\in -\Delta^1$, and by the monotonicity of $\TD$, $q\in
-\Delta_{F}$.   
\end{case}

\begin{case}[Inductive step]{}
  Let us assume that the property holds up to $n$ and $P(n+1)=-\Delta
  p$. This implies 1) $p\notin F$ and 2) $\forall r\in R_s[p]\exists
  q\in A(r)$ such that $-\Delta q\in P (1..n)$.  By inductive
  hypothesis, for some $m,$ $q\in-\Delta^m$ and so
  $A(r)\cap-\Delta^m\neq\emptyset$, thus $p\in=\Delta^{m+1}$ and
  therefore by the monotonicity of $\TD$ it is in $-\Delta_F$.
\end{case}

\begin{case}[Inductive base $\Leftarrow$]{}
 As we have seen $-\Delta^1=\{p|\neg\exists r\in R_s \wedge p\notin
 F\}$, thus, vacuously, we have a single line proof of $-\Delta q$.  
\end{case}

\begin{case}[Inductive step]{}
  Let us assume that the property holds up to $n$, and let us suppose
  that $q\notin -\Delta^n$, but $q\in -\Delta^{n+1}$. This implies
  that $q\notin F$ and $\forall r\in R_s[q]:
  A(r)\cap-\Delta^n\neq\emptyset$. This means that $\forall r\in
  R_s[q]$ $\exists a_r\in A(r)$ such that $D\vdash -\Delta a_r$. Let
  $P(a)$ be a derivation of $a$, and $P(a_1),\dots,P(a_r)$ be the
  concatenation of the proofs of such $a$s. We append $-\Delta q$ at
  the end and we obtain a proof of $-\Delta q$, thus $D\vdash-\Delta
  q$.
\end{case} 

\begin{case}[\textbf{Cases $+\partial$ and $-\partial$}. Inductive base, ]{$\Rightarrow$}
  The tags $+\partial$ and $-\partial$, as well as the corresponding
  sets of literals, depend on each other, so we will carry out the
  proofs simultaneously. Moreover, since the tags $+\Delta$ and
  $-\Delta$ (and the sets of literal corresponding to them) are
  independent from $+\partial$ and $-\partial$ --and we have already
  proved the theorem for them-- we assume, without any loss of
  generality, that derivations in defeasible logic consist only of
  defeasible tagged literals.
\end{case}

\begin{case}{$P(1)=+\partial q$} 
This is possible if 1) $q\in +\Delta_F$, or  2) $\non q\in-\Delta_F$
and $\exists r\in R_{sd}[q]$ such that $A(r)=\emptyset$ and there are
no rules for $\non p$. For 1) by the definition of $\TD$ there exists
an $n$ such that $q\in +\Delta^n$, then $q\in +\partial^{n+1}$; by the
monotonicity of $\TD$ $q\in+\partial_F$. For 2) we have that $A(r)$ is
included in every $+\partial^n$, and the condition beginning with $\forall r\in
R[\non q]$ is vacuously satisfied since $R[\non q]=\emptyset$. Let
$-\Delta^m$ be the minimal set in the Kleene sequence generated by $\TD$  
such that $\non q\in-\Delta^m$. Then we can conclude that $q\in
+\partial^{n+1}$; therefore $q\in+\partial_F$.
\end{case}

\begin{case}{$P(1)=-\partial q$}
First of all, let us consider the following decomposition of
$-\partial^{n+1}$
\[
-\partial^{n+1}=-\partial^{n+1}_{\cap} \cup
                -\partial^{n+1}_{\Delta} \cup
                -\partial^{n+1}_{>}
\]
where
\begin{align*}
  -\partial^{n+1}_{\cap} &= \{p \in -\Delta^n| \forall r\in
     R_{sd}[p],A(r)\cap -\partial^n \neq\emptyset\}\\
  -\partial^{n+1}_{\Delta} &= \{p\in-\Delta^n|\non p\in +\Delta^n\}\\
  -\partial^{n+1}_{>} &= \{p\in -\Delta^n|\forall r\in R_{sd}[p] 
  \begin{array}[t]{l}
    \exists s\in R[\non p]: A(s)\subseteq +\partial^n \text{ and}\\
    \forall t\in R_{sd}[p]: \text{ either }A(t)\cap
      -\partial^n\neq\emptyset \text{ or } t\not>s\}
  \end{array}
\end{align*}
Now $P(1)=-\partial q$ is possible if $D\vdash -\Delta q$ and either
\begin{enumerate}[~~~~~]
\item $D\vdash +\Delta q$ or
\item $R_{sd}[\non q]=\emptyset$ or
\item $\exists s\in R[\non q]: A(s)=\emptyset$ and $\neg \exists t\in
  R[q]: t> s$
\end{enumerate}
From the previous cases we have $q\in -\Delta_F$ and for 1) $\non q\in
+\Delta_F$. Let $n$ be the minimum number of iterations of $\TD$ such
that both $q\in -\Delta^n$ and $q\in +\Delta^n$; then by construction
$q\in +\partial^{n+1}$. For 2) we obtain that $q\in
-\partial^{n+1}_{\cap}$ and $q\in -\partial^{n+1}_{>}$ since the
conditions are vacuously satisfied. Finally 3) implies that
$A(s)\subseteq +\partial^n$ for any $n$ and $\forall t\in R[q], t\not>
s$; thus $q\in-\partial^{n+1}_{>}$ for some $n$ such that $q\in
+\Delta^n$.
\end{case}

\begin{case}[Inductive step]{}\ 
\begin{case}{$P(n+1)=+\partial q$}
  Let us assume that the inductive hypothesis holds for derivations
  of length up to $n$.  We only show the cases different from the
  inductive base. This means that we have cases corresponding to
  clause 2) of the proof conditions for $+\partial$. From the
  inductive hypothesis we have $A(r)\subseteq +\partial_F$ for some
  $r\in R_{sd}[q]$ (clause 2.2), $\forall s\in R[\non q]$ either
  $A(s)\cap -\partial_F\neq\emptyset$ (clause 2.3.1) or $\exists t\in
  R_{sd}[q]: t>s$ and $A(t)\subseteq +\partial_F$ (clause 2.3.2). By
  the monotonicity of $\TD$ we get that a minimum $n$ such that
  $+\partial^n$ and $-\partial^n$ that satisfy the above condition
  exists. Therefore $q\in +\partial^{n+1}$ and consequently $q\in
  +\partial_F$. 
\end{case}

\begin{case}{$P(n+1)=-\partial q$}
  Here we have 1) $q\in -\Delta_F$, and either 2) $\non q\in+\Delta_F$
  or 3) the clause 2.1 of the proof condition for $-\partial$ is not
  vacuously satisfied: in such a case $\forall r\in R_{sd}[p]$:
\begin{enumerate}[~~~~~~]
\item[2.1] $\exists p\in A(r)$ such that $-\partial q\in P(1..n)$;
  by inductive hypothesis $p\in-\partial_{F}$, so for some $m$,
  $A(r)\cap-\partial^n\neq\emptyset$.
\item[2.3] $\exists s\in R[\non q]$ such that 
  \begin{itemize}
  \item[2.3.1] $\forall a_i\in A(s) +\partial a_{i}\in P(1..n)$;
    by inductive hypothesis each $q_i\in+\partial_F$, therefore for
    some $m$, $A(s)\subseteq+\partial^m$; or
  \item[2.3.2] $\forall t\in R[q]$ either $t\not>s$, or $\exists a\in
    A(t)$ such that $-\partial q\in P(1..n)$. By inductive
    hypothesis $a\in-\partial_F$, then, for some $m$
    $A(t)\cap-\partial^m\neq\emptyset$.
  \end{itemize}
\end{enumerate}
Also in this case it is immediate to see that each condition has been
reduced to the correspondent condition required for the construction
of $-\partial^{m+1}$, therefore, for the smallest $m$ satisfying the
above three conditions we can conclude $q\in-\partial^{m+1}$, and, by
the monotonicity of $\TD$, $q\in-\partial_F$.
\end{case}
\end{case}
%%%%%%%%%%%%%%%%%%%%%%%%%%%%%%%%%%%%%%%%%%%%%%%%%%
% Inductive proof <= +\partial -\partial
%%%%%%%%%%%%%%%%%%%%%%%%%%%%%%%%%%%%%%%%%%%%%%%%%%
\begin{case}[Inductive base, $\Leftarrow$]{}
\begin{case}{$q\in +\partial^1$}
If $q\in +\partial^1$ because it is in $+\Delta_F$, then we have that
$D\vdash +\Delta q$. Let $P$ be a proof of $+\Delta q$; we append
$+\partial q$ at the end of $P$ obtaining a proof of $+\partial q$.

For the other case we have 
\[
\exists r\in R_{sd}[q]: A(r)=\emptyset
\text{ and }
\forall s\in R[\non q]\ \exists t\in R[q]: A(t)=\emptyset, t>s
\]
In this case it is easy to verify that $+\partial q$ is a single line
proof for $q$.
\end{case}

\begin{case}{$q\in -\partial^1$}
Here we have that $q\in -\Delta_F$ and the following three cases:
\begin{enumerate}[~~~~~]
\item $\non q \in +\Delta_F$
\item $R_{sd}[q]=\emptyset$
\item $\exists r\in R[\non q]: A(r)=\emptyset$ and $\forall t\in R[q],
  t\not>r$ 
\end{enumerate}
In the first case, we have already proved that $D\vdash+\Delta q$ and
$D\vdash-\Delta q$. So let $P$ the concatenation of the proofs of
$+\Delta q$ and $-\Delta q$. We append $-\partial q$ at the end of $P$
and we obtain a derivation of $-\partial q$.  For 2) the sequence
$-\Delta q, -\partial q$ is vacuously a derivation of $-\partial q$,
and for 3) the concatenation of a proof of $-\Delta q$ and $-\partial
q$ satisfies condition 3 of the definition of $-\partial$.
\end{case}

\begin{case}[Inductive step]{}
  We assume that the theorem holds up to $+\partial^n$ and
  $-\partial^n$.  Furthermore it is worth noting that the construction
  of derivation of defeasible literals in Defeasible Logic and the
  construction of the corresponding sets of literals are invertible,
  i.e., each time the top-down (procedural) condition requires
  something to be provable, the bottom-up (set-theoretic) condition
  satisfies an appropriate set-theoretic condition. But we have seen
  that, granted the inductive hypothesis the proof-theoretic condition
  and the set-theoretic one coincide.
\end{case}

\begin{case}{$q\in +\partial^{n+1}$}
The difference with the inductive base is that now we replace
$A(r)=\emptyset$, $A(t)=\emptyset$ with $A(r)\subseteq +\partial^n$,
$A(t)\subseteq +\partial^n$, and $A(s)\cap
-\partial^n\neq\emptyset$. By the inductive hypothesis we have
$\forall a_r\in A(r): D\vdash +\partial a_r$,  $\forall a_t\in A(t):
D\vdash +\partial a_t$, and $\exists a_s\in A(s): D\vdash -\partial
a_s$. Let $P$ be a concatenation of the derivations of the just
mentioned tagged literals. We append $+\partial q$ at the end of $P$
and we obtain a derivation of $+\partial q$.
\end{case}

\begin{case}{$q\in -\partial^{n+1}$}
  First of all we can use an argument similar to the previous case.
  Here the main difference with the inductive base is that we have to
  consider that the set of supportive rules for $q$ is not empty and
  we have to consider the case where $\forall r\in R_{sd}[q], A(r)\cap
  -\partial^n\neq \emptyset$. By inductive hypothesis we have that
  $\forall r\in R_{sd}[q]\exists a_r\in A(r): D\vdash -\partial a_r$.
  At this stage we can concatenate the derivations of such $a_r$s with
  the other required derivations and we append $-\partial q$ at the
  end. Again we can verify that the resulting sequence is a proof for
  $-\partial q$.
\end{case}
\end{case}
\end{proof}

\subsection{Beyond Propositional Defeasible Logic}
\label{sec:beyond-prop-defe}

The restriction of Theorem \ref{butd} to finite propositional theories
derives from the formulation of the proof theory; proofs are
guaranteed to be finite under this restriction.  Defeasible Logic has
a constructive proof theory that guarantees the explainability of
conclusions (i.e., for every conclusion we derive, it is possible to
provide a full proof with justification of the essential steps). In
addition the proof theory defines the classical defeasible logic of
\cite{Antoniou00c} and \cite{Billington90}.  

On the other hand, the bottom-up semantics does not need this
restriction, and so can be used in conjunction with predicate
defeasible logic rules that represent infinitely many propositional
rules. This also means that under this characterisation defeasible
logic can be supplemented with function symbols. In the following we
will take advantage of this opportunity and provide a more elegant and
general characterisation under the Kunen semantics (Section 3.2 and
5.3), while in the remaining sections we stick to the original
definition.

\section{Semantics of Logic Programs}

Now that we have presented defeasible logic and have developed the
technical properties that will be needed later on, we turn to the
theme of this paper: establish formal connections between defeasible
logic and logic programming semantics. This section presents the
basics of the semantics that will be used in subsequent sections

A {\em logic program} $P$ is a finite set of program clauses. A {\em
  program clause} $r$ has the form
\[
A \leftarrow B_1,\ldots,B_n,\mathit{not} \ C_1,\ldots,\mathit{not} \ C_m
\]
where $A,B_1,\ldots B_n,C_1,\ldots,C_m$ are positive literals. A
program clause with variables is considered to represent the set
of its ground instances.

In this paper we will make use of two well-known logic programming
semantics: stable model semantics \cite{GL88} and Kunen semantics
\cite{Kunen}. In the following we present them briefly for the sake of
completeness.

\subsection{Stable Model Semantics}

Let $M$ be a subset of the Herbrand base. We call a ground program
clause
\[
A\leftarrow B_1,\ldots,B_n,\mathit{not} \ C_1,\ldots,\mathit{not} \ C_m
\]
\emph{irrelevant w.r.t.\ $M$} if at least one $C_i$ is included in $M$.
Given a logic program $P$, $ground(P)$ is the set of ground instances
of the logic program $P$, and we define the {\em reduct of $P$ w.r.t.\ 
  $M$}, denoted by $P^M$, to be the logic program obtained from
$ground(P)$ by 

\begin{enumerate}

\item removing all clauses that are irrelevant w.r.t.\ $M$, and

\item removing all premises $not \ C_i$ from all remaining program
clauses.

\end{enumerate}
Note that the reduct $P^M$ is a definite logic program, and we are
no longer faced with the problem of assigning semantics to
negation, but can use the least Herbrand model instead.

$M$ is a {\em stable model} of $P$ iff $M={\cal M}_{P^M}$, where
$\mathcal{M}_{P^M}$ is the least Herbrand model of the reduct of $P$.

\subsection{Kunen Semantics}

Kunen semantics \cite{Kunen} is a 3-valued semantics for logic
programs. A {\em partial interpretation} is a mapping from ground
atoms to one of the three truth values {\bf t}, {\bf f} and {\bf u},
which denote true, false and unknown, respectively. This mapping can
be extended to arbitrary formulas using Kleene's 3-valued logic.

Kleene's truth tables can be summarized as follows. If $\varphi$ is a
boolean combination of atoms with truth value one of {\bf t}, {\bf f}
and {\bf u}, its truth value is {\bf t} iff all possible ways of
putting {\bf t} or {\bf f} for the various {\bf u}-values lead to a
value {\bf t} being computed in ordinary (2-valued) logic; $\varphi$
gets the value {\bf f} iff $\neg \varphi$ gets the value {\bf t}; and
$\varphi$ gets the value {\bf u} otherwise.  These truth values can be
extended in the obvious way to predicate logic, thinking of the
quantifiers as infinite conjunctions or disjunctions.

The Kunen semantics of a program $P$ is obtained from a sequence
$\{I_n\}$ of partial interpretations, defined as follows:

\begin{enumerate}

\item $I_0(a) = $ {\bf u} for every atom $a$.

\item $I_{n+1}(a) = $ {\bf t} iff there is a program clause
\[
A \leftarrow B_1,\ldots,B_n,not \ C_1,\ldots,not \ C_m
\]
and a ground substitution $\sigma$ such that $a=A\sigma$ and that
\[
I_n((B_1\wedge\ldots\wedge B_n\wedge \neg C_1\wedge
\ldots\wedge \neg C_m)\sigma) = {\bf t}.
\]
\item $I_{n+1}(a) = $ {\bf f} iff for all clauses
\[
A \leftarrow B_1,\ldots,B_n,not \ C_1,\ldots,not \ C_m
\]
in the program, and all ground substitutions $\sigma$, if  $a =
A\sigma$ then
\[
I_n((B_1\wedge\ldots\wedge B_n\wedge \neg
C_1\wedge \ldots\wedge \neg C_m)\sigma) =  {\bf f}.
\]
\item $I_{n+1}(a) = $ {\bf u} if neither 2. nor 3. applies.
\end{enumerate}
We shall say that the Kunen semantics of $P$ supports a formula
$\varphi$, written $P\models_K \varphi$, iff there is an
interpretation $I_n$, for some finite $n$, such that $I_n(\varphi)
= $ {\bf t}.

\section{A Translation of Defeasible Theories into Logic Programs}

In this section we describe a meta-program \M in a logic
programming form that expresses the essence of the defeasible
reasoning embedded in defeasible logic first introduced in
\cite{Maher99}. \M consists of the following clauses. We first
introduce the predicates defining classes of rules, namely
\begin{clause}
$\mt{supportive\_rule}(Name,Head,Body)$:-\\
  \> $\mt{strict}(Name,Head,Body)$.\\
\\[-.5\baselineskip]
$\mt{supportive\_rule}(Name,Head,Body)$:-\\
  \> $\mt{defeasible}(Name,Head,Body)$.
\end{clause}

\begin{clause}
  $\mt{rule}(Name,Head,Body)$:-\\
  \> $\mt{supportive\_rule}(Name,Head,Body)$.\\
\\[-.5\baselineskip]
  $\mt{rule}(Name,Head,Body)$:-\\
  \> $\mt{defeater}(Name,Head,Body)$.
\end{clause}
Next we introduce the clauses defining the predicates corresponding to
$\PD{}$, $\MD{}$, $\pd{}$, and $\md{}$. These clauses specify the
structure of defeasible reasoning in defeasible logic. Arguably they
convey the conceptual simplicity of defeasible logic more clearly than
the proof theory.
\begin{Clause}\label{strictly1}
  $\mt{definitely}(X)$:-\\
  \> $\mt{fact}(X)$.
\end{Clause}

\begin{Clause}\label{strictly2}
  $\mt{definitely}(X)$:-\\
  \> $\mt{strict}(R,X,[\seq{Y}])$,\\
  \> $\mt{definitely}(Y_1)$,\dots,$\mt{definitely}(Y_n)$.
\end{Clause}

\begin{Clause}\label{defeasibly1}
  $\mt{defeasibly}(X)$:-\\
  \> $\mt{definitely}(X)$.
\end{Clause}

\begin{Clause}\label{defeasibly2}
  $\mt{defeasibly}(X)$:-\\
  \> $\mt{not\ definitely}(\sim X)$,\\
  \> $\mt{supportive\_rule}(R,X,[\seq{Y}])$,\\
  \> $\mt{defeasibly}(Y_1)$,\dots,$\mt{defeasibly}(Y_n)$,\\
  \> $\mt{not\ overruled}(R,X)$.
\end{Clause}

\begin{Clause}\label{overruled}
  $\mt{overruled}(R,X)$:-\\
%  \> $\mt{sup}(S,R)$,\\
  \> $\mt{rule}(S,\sim X,[\seq{U}])$,\\
  \> $\mt{defeasibly}(U_1)$,\dots,$\mt{defeasibly}(U_n)$,\\
  \> $\mt{not\ defeated}(S,\sim X)$.
\end{Clause}

\begin{Clause}\label{defeated}
  $\mt{defeated}(S,\sim X)$:-\\
  \> $\mt{sup}(T,S)$, \\
  \> $\mt{supportive\_rule}(T,X,[\seq{V}])$,\\
  \> $\mt{defeasibly}(V_1)$,\dots,$\mt{defeasibly}(V_n)$.
\end{Clause}

The first two clauses address definite provability, while the
remainder address defeasible provability. The clauses specify if
and how a rule can be overridden by another, and which rules can
be used to defeat an overriding rule, among other aspects of the
structure of defeasible reasoning in defeasible logic.

We have permitted ourselves some syntactic flexibility in
presenting the meta-program. However, there is no technical
difficulty in using conventional logic programming syntax to
represent this program.

Finally, for a defeasible theory $D=(F,R,>)$, we add facts
according to the following guidelines:
\begin{enumerate}
\item $\mt{fact}(p)$.  \hfill for each $p\in F$
\item $\mt{strict}(r_i,p,[q_1,\dots,q_n])$.
\hspace*{\fill} for each rule
  $r_i:q_1,\dots,q_n\to p\in R$
\item $\mt{defeasible}(r_i,p,[q_1,\dots,q_n])$.
\hspace*{\fill} for each rule
  $r_i:q_1,\dots,q_n\Rightarrow p\in R$
\item $\mt{defeater}(r_i,p,[q_1,\dots,q_n])$.
\hspace*{\fill} for each rule
  $r_i:q_1,\dots,q_n\leadsto p\in R$
\item $\mt{sup}(r_i,r_j)$.
\hspace*{\fill} for each pair of rules such that
  $r_i > r_j$
\end{enumerate}

\section{Properties of the Translation}

\subsection{Embedding under Stable Model Semantics}

We establish relationships between $D$ and its translation \M. To do
so we must select appropriate logic program semantics to interpret
$not$ in clauses \ref{defeasibly2} and \ref{overruled}.  First we
consider stable model semantics. 

The aim of this section is twofold. On one hand it established a
relationship between Defeasible Logic and stable semantics. This
connection is obtained via the representation of the meta-program for
Defeasible Logic as a Default Theory. In this way we are able to use
the well-know and well-understood link between stable semantics and
Default Logic to simplify the proofs of our results and, at the same
time, it opens the way to investigations on the similarities and
differences over the two non-monotonic formalisms. Furthermore it will
enable further studies on relationships between semantics for default
logic and defeasible logic.

%Stable model semantics for the meta-program will be obtained by its
%representation in default logic, making use of a well-known connection
%between stable model semantics and default logic. 
To this end we
briefly rehearse the basic definitions of default logic.

\subsubsection{Basics of Default Logic}

A {\em default} $\delta$ has the form
${\varphi:\psi_1,\ldots,\psi_n\over\chi}$ with closed formulas
$\varphi,$ $\psi_1,\ldots$,$ \psi_n,$ $\chi$. $\varphi$ is the
{\em prerequisite} $pre(\delta)$, $\psi_1,\ldots,\psi_n$ the {\em
justifications} $just(\delta)$, and $\chi$ the {\em consequent}
$cons(\delta)$ of $\delta$.

A {\em default theory} $T$ is a pair $(W,Def)$ consisting of a set
of formulas $W$ and a countable set $Def$ of defaults.

Let $\delta = {\varphi:\psi_1,\ldots,\psi_n\over\chi}$ be a
default, and $E$ a deductively closed set of formulas. We say that
$\delta$ {\em is applicable to} $E$ iff $\varphi\in E$, and $\neg
\psi_1,\ldots,\neg\psi_n\not\in E$.

Let $\Pi=(\delta_0,\ \delta_1,\ \delta_2,\ldots)$ be a finite or
infinite sequence of defaults from $Def$ without multiple
occurrences (modelling an application order of defaults from
$Def$). We denote by $\Pi[k]$ the initial segment of $\Pi$ of
length $k$, provided the length of $\Pi$ is at least $k$.

\begin{itemize}
\item $In(\Pi) = Th(W\cup\{cons(\delta) \ | \ \delta$ occurs in
$\Pi\})$, where $Th$ denotes the deductive closure.

\item $Out(\Pi) = \{\neg\psi \ | \ \psi\in just(\delta), \delta$
occurs in $\Pi\}$.

\end{itemize}
$\Pi$ is called a {\em process of} $T$ iff $\delta_k$ is
applicable to $In(\Pi[k])$, for every $k$ such that $\delta_k$
occurs in $\Pi$. $\Pi$ is {\em successful} iff $In(\Pi)\cap
Out(\Pi)= \emptyset$, otherwise it is {\em failed}. $\Pi$ is {\em
closed} iff every default that is applicable to $In(\Pi)$ already
occurs in $\Pi$.

\citeN{Antoniou97} showed that Reiter's \citeyear{Reiter80} original
definition of extensions is equivalent to the following one: A set of
formulas $E$ is an {\em extension} of a default theory $T$ iff there
is a closed and successful process $\Pi$ of $T$ such that $E=In(\Pi)$.

\subsubsection{Stable Models and Default Logic}

The Default Logic interpretation of a ground program clause $Cl$

$$A\leftarrow B_1,\ldots,B_n,not \ C_1,\ldots,not \ C_m$$
is given by the default

$$df(Cl) = {B_1\wedge\ldots\wedge B_n: \neg \ C_1,\ldots,\neg \ C_m\over
A}.$$ We define $df(P)$, the {\em default logic interpretation} of
the logic program $P$, to be the default theory $(W,D)$ with
$W=\emptyset$ and $D=\{df(Cl)\ | \ Cl\in ground(P)\}$.

\begin{theorem}[\citeNP{Antoniou97}]
  Let $P$ be a logic program, and $M$ a subset of the Herbrand
  base. $M$ is a stable model of $P$ iff $Th(M)$ is an extension of
  $df(P)$.
\end{theorem}

\subsection{The Meta-Program as a Default Theory}

According to the previous section, the meta-program of section 4
can be equivalently viewed as the following default theory
$T(D)=(W(D),Def(D))$.

The predicate logic part $W(D)$  contains:

\begin{enumerate}

\item $\fact(p)$ \hfill for each $p\in F$

\item $\mt{strict}(r_i,p,[q_1,\dots,q_n])$.
\hspace*{\fill} for each rule
  $r_i:q_1,\dots,q_n\to p\in R$
\item $\mt{defeasible}(r_i,p,[q_1,\dots,q_n])$.
\hspace*{\fill} for each rule
  $r_i:q_1,\dots,q_n\Rightarrow p\in R$
\item $\mt{defeater}(r_i,p,[q_1,\dots,q_n])$.
\hspace*{\fill} for each rule
  $r_i:q_1,\dots,q_n\leadsto p\in R$
\item $\mt{sup}(r_i,r_j)$.
\hspace*{\fill} for each pair of rules such that
  $r_i > r_j$

\end{enumerate}
And the following defaults (actually default schemas) are elements of
$Def(D)$:\\

\noindent $d_1= {\fact(X): \over \definitely(X)}$ \\

\noindent $d_2= {\strict(R,X,[Y_1,\ldots,Y_n])\wedge
\definitely(Y_1)\wedge\ldots\wedge \definitely(Y_n): \over
\definitely(X)}$ \\

\noindent $d_3= {\definitely(X): \over \defeasibly(X)}$ \\

\noindent $d_4= {
{\begin{array}{r}
\multicolumn{1}{l}{\scriptstyle\suprule(R,X,[Y_1,\ldots,Y_n])\wedge
\defeasibly(Y_1)\wedge\ldots\wedge \defeasibly(Y_n):\qquad\qquad}\\
\scriptstyle\neg
\definitely(\non X), \neg \overruled(R,X) 
\end{array}}
\over \defeasibly(X)}$ \\

\noindent $d_5= {\drule(S,\non X,[U_1,\ldots,U_n])\wedge
\defeasibly(U_1)\wedge\ldots\wedge \defeasibly(U_n):\neg
\defeated(S,\non X) \over \overruled(R,X)}$ \\

\noindent $d_6= {\mt{sup}(T,S)\wedge
\suprule(T,X,[V_1,\ldots,V_n])\wedge
\defeasibly(V_1)\wedge\ldots\wedge \defeasibly(V_n): \over
\defeated(S,X)}$ \\

\noindent $d_7= {strict(Name,Head,Body):\over
\suprule(Name,Head,Body)}$ \\

\noindent $d_8= {defeasible(Name,Head,Body):\over
\suprule(Name,Head,Body)}$ \\

\noindent $d_9= {\suprule(Name,Head,Body):\over
\drule(Name,Head,Body)}$ \\

\noindent $d_{10}= {\mt{defeater}(Name,Head,Body):\over
\drule(Name,Head,Body)}$ \\

Now let us prove a technical result on this default theory. It
provides a  condition on $D$ under which $T(D)$ has at least one
extension.

\begin{lemma}
Let $D$ be a decisive defeasible theory. Then $T(D)$ has at least
one extension.
\end{lemma}

\begin{proof}

If the atom dependency graph of $D$ is acyclic we can define an
arbitrary total order $\gg$ on atoms which respects the dependency
graph. We proceed to construct a closed and successful process
$\Pi$ as follows:

\begin{enumerate}[~~~~~]

\item First apply instantiations of $d_1$ and $d_2$ in any order. These
defaults have no justification, so success cannot be jeopardized.

\item Then proceed to prove $\defeasibly(p)$ or $\defeasibly(\neg p)$ using the
remaining defaults, in the order of $\gg$. 

\item For each atom $p$, try to apply first defaults $d_3$, then
$d_6$, then $d_5$ and finally $d_4$. If the defaults can be
applied in this order only, the process $\Pi$ cannot fail.

\item $\Pi$ is closed when we have carried out step 3 for all
atoms $p$.

\end{enumerate}
The question is whether the order specified in 3 can always be
respected. This is the case because $Y_i$, $U_i$ and $V_i$ use
atoms appearing before $X$ in the total order.

The argument in more detail: we analyze the situation where an
instantiation of $d_4$ with $p$ has impact on an earlier
application of $d_5$ with instantiation $\non q$. Suppose
$\defeasibly(p)$ is derived using default $d_4$ at stage $k$, and
suppose it is then used to prove $\overruled(r, q)$ later in the
process $\Pi$, where $\defeasibly(q)$ was derived in $\Pi$ at a
stage $l<k$. But then $p$ occurs in the body of a rule with head
predicate $q$, so $q$ depends on $p$, so it must appear {\em
after} $p$ in $\gg$. So, according to 2 above, $d_4$ with
instantiation $q$ cannot have been applied before $d_4$ with
instantiation $p$, so we have a contradiction.

A similar argument applies to the interplay between defaults $d_5$
and $d_6$.
\end{proof}

In general, $T(D)$ may not have any extension if the condition of
the Lemma is not satisfied. For example consider $D$ consisting of
the rules:

\begin{quote}
$r_1: \ \Rightarrow p$

$r_2: \ p \Rightarrow q$

$r_3: \ q \Rightarrow \neg p$
\end{quote}
Let us now analyze the application of defaults in $T(D)$.

Because of $r_1$ we derive immediately $\overruled(r_3)$
using $d_5$ with instantiation $r_3$ for $R$ and $r_1$ for $S$.
$r_1$ cannot be defeated using $d_6$ because there is no stronger
rule. No interaction with other defaults can prevent this
application of $d_5$, so it can appear at the beginning of any
process, without loss of generality.

Then we can apply $d_4$ to derive $\defeasibly(p)$, assuming $\neg
\overruled(r_1,p) \ (*)$.

Now that $\defeasibly(p)$ is derived, we can apply $d_4$ to derive
$\defeasibly(q)$, assuming $\neg \overruled(r_2,q)$.

We apply $d_5$ with instantiation $r_3$ for $S$ and $r_1$
for $R$ to derive $\overruled(r_1,p)$. This contradicts the
previous assumption $(*)$, so the process is failed.

There is no way any defaults along the process above can be
blocked by applying another default instead. So there can be no
extension.

We are now able to prove the main results, namely the relationships
between Defeasible Logic and the stable semantics interpretation of
the meta-program describing provability in Defeasible Logic.

\begin{theorem} \label{stable-definitely}
\begin{itemize}

\item[(a)] If $D\vdash +\Delta p$ then  $\definitely(p)$ is included in all
stable models of \M.
\item[(b)] If $D\vdash -\Delta p$ then $\definitely(p)$ is not included in
any stable model of \M.
\item[(c)] If $D$ is decisive then
the implications (a) and (b) are also true in the opposite
direction.

\end{itemize}

\end{theorem}
\begin{proof}

(a): Proof by induction on the length of derivations $P$ in $D$.
Let the claim hold for $P(1..i)$, and let $P(i+1) = +\Delta p$.
Let $E$ be an extension of $T(D)$, and $E=In(\Pi)$ for a closed
and successful process $\Pi$ of $T(D)$.

{\em Case 1}: $p\in F$. Then $\fact(p)\in W(D)$. Since $d_1$ is
applicable to $In(\Pi)$ and $\Pi$ is closed, we conclude
$\definitely(p) \in In(\Pi)=E$.

{\em Case 2}: There is $r\in R_s[p]$ such that $+\Delta a\in
P(1..i)$ for all $a\in A(r)$. Since $r\in R_s[p]$ we have
$strict(r,p,[q_1,\ldots,q_n]) \in Def(D)$. Since $+\Delta q_j \in
P(1..i)$, for all $j=1,\ldots, n$, we conclude with Induction
Hypothesis that $\definitely(q_j) \in E = In(\Pi)$. Thus $d_2$ is
applicable to $In(\Pi)$. $\Pi$ is closed so $\definitely(p)\in
In(\Pi) = E$.

(b): The proof goes by induction on the length of derivations $P$
in $D$. Let the claim hold for $P(1..i)$ and let $P(i+1) = -\Delta
p$. Further let $E$ be an extension of $T(D)$. Then $E=In(\Pi)$
for a closed and successful process $\Pi$ of $T(D)$.

By the inference condition $(-\Delta)$ we know $p\not\in F$, thus
$fact(p) \not\in W(D)$. $(*)$

Also we know $\forall r\in R_s[p] \exists a\in A(r): -\Delta a\in
P(1..i)$. By induction Hypothesis we conclude

\[
\forall r\in R_s[p] \exists a\in A(r): \definitely(a) \not\in
In(\Pi)\tag{**}
\]
$(*)$ and $(**)$ show that neither $d_1$ nor $d_2$ can be used to
derive $\definitely(p)$ in $\Pi$. But these are, by construction of
$T(D)$, the only possibilities. Thus $definitely(p)\not\in
In(\Pi)=E$.

(c): Let $\definitely(p)\in E$ for an extension $E$ of $T(D)$. Such
an extension exists because $D$ is decisive. Then, by part (b) we
conclude $D\not\vdash -\Delta p$. Therefore $D\vdash +\Delta p$
because $D$ is decisive. Thus the opposite of (a) holds. The
opposite of (b) is shown in an analogous way.

Consider $D=\{{}\Rightarrow p, \ p\Rightarrow q, \ q\Rightarrow \neg
p\}$. As we have shown in section A.3, $T(D)$ has no extension, so
$\definitely(p)$ is included in all extensions of $T(D)$. However
$D\not\vdash +\Delta p$. Thus the opposite of (a) does not hold,
in general.

Consider the theory $D$ consisting only of the strict rule
$p\rightarrow p$. $T(D)$ has only one extension,
$E=Th(\emptyset)$, and $\definitely(p)\not\in E$. However
$D\not\vdash -\Delta p$. This shows that the opposite of (b) is
not true, in general.
\end{proof}

\begin{theorem} \label{stable-defeasibly}
\begin{itemize}
\item[(a)] If $D\vdash +\partial p$ then $\defeasibly(p)$ is
  included in all stable models of \M.
\item[(b)] If $D\vdash -\partial p$ then $\defeasibly(p)$ is not
  included in any stable model of \M.
\item[(c)] If $D$ is decisive then the implications (a) and (b) are
  also true in the opposite direction.
\end{itemize}

\end{theorem}
\begin{proof}

(c): Let $\defeasibly(p)\in E$ for an extension $E$ of $T(D)$. Such
an extension exists because $D$ is decisive. Then, by part (b) we
conclude $D\not\vdash -\partial p$. Therefore $D\vdash +\partial
p$ because $D$ is decisive. Thus the opposite of (a) holds. The
opposite of (b) is shown in an analogous way.

Let $D$ consist of the rules: $p\Rightarrow p$, $p\Rightarrow q$
and $\Rightarrow \neg q$. Then $\defeasibly(\neg q)$ is included in
the only extension of $T(D)$ but $D\not\vdash \neg q$. This shows
that the opposite of (a) is not necessarily true if $D$ is not
decisive.

A counterexample for the opposite direction of (b) is the
defeasible theory consisting only of the rule $p\Rightarrow p$.
$\defeasibly(p)$ is not included in the only extension of $T(D)$,
however $-\partial p$ cannot be derived from $D$.

\vspace{4pt} Parts (a) and (b) are shown concurrently by induction
on the length of a derivation $P$ in $D$. Suppose (a) and (b) hold
for $P(1..i)$ (Induction Hypothesis). Consider an extension $E$ of
$T(D)$, and let $E=In(\Pi)$ for a closed and successful process
$\Pi$ of $T(D)$.

\vspace{4pt} {\bf Case $+\partial$}: $P(i+1) = +\partial
p$. By the inference condition $(+\partial)$ there are two cases.
The first case is that $+\Delta p\in P(1..i)$. By Theorem 5.2,
$\definitely(p)\in E=In(\Pi)$. Then the default $d_3$ (with
instantiation $p$ for $X$) is applicable to $In(\Pi)$. Since $\Pi$
is closed we conclude $\defeasibly(p)\in In(\Pi) = E$.

The second case is as follows:

\begin{tabbing}
\=7890\=1234\=5678\=9012\=3456\=\kill
 \>\>(1) $\exists
r\in R_{sd}[q]\ \forall a \in
A(r): +\partial a\in P(1..i)$ and \\
\>\>(2) $-\Delta\non q \in P(1..i)$ and \\
\>\>(3) $\forall s \in R[\non q]$ either \\
\>\>\>(3.1) $\exists a\in A(s): -\partial a\in P(1..i)$ or \\
\>\>\>(3.2) $\exists t\in R_{sd}[q]$ such that \\
\>\>\>\> $\forall a\in A(t): +\partial a\in P(1..i)$ and $t>s$
\end{tabbing}
From (2) we conclude
\[
\definitely(\non p)\not\in In(\Pi)=E \tag{*}
\]
using Theorem A.2. From (1) we get
\[
\suprule(r,p,[q_1,\ldots,q_n]) \in W(D) \tag{**}
\]
From (1) and Induction Hypothesis we get
\[
\defeasibly(q_i) \in In(\Pi), \mbox{ for all }i=1,\ldots,n \tag{***}
\]
In the following we show that $\overruled(r,p)\not\in In(\Pi)$.
Together with $(*)$-$(***)$ it shows that the default $d_4$ (with
instantiation $p$ for $X$) is applicable to $In(\Pi)$. Since $\Pi$
is closed we get $\defeasibly(p) \in In(\Pi) = E$.

Consider $s\in R[\non p]$. In case (3.1) holds we have $-\partial
a\in P(1..i)$ for an $a\in A(s)$. By Induction Hypothesis we
conclude $\defeasibly(a)\not\in In(\Pi)=E$. Thus default $d_5$
cannot be applied with $s$ instantiated for $S$.

In case (3.2) holds, it is easily seen that default $d_6$ can be used
for the derivation of $\defeated(s,\non p)$.  Thus $\defeated(s,\non
p) \in In(\Pi)$, and $d_5$ cannot be applied with $s$ instantiated for
$S$.

Overall we have shown that $d_5$ fails to derive $\overruled(r,p)$.

\vspace{4pt} {\bf Case $-\partial$}: Let $P(i+1) =
-\partial p$. From the $(-\partial)$ inference condition we know
$-\Delta\in P(1..i)$. Therefore, by Theorem A.2,
$\definitely(p)\not\in E=In(\Pi)$. Thus default $d_3$ cannot be
used to derive $\defeasibly(p)$ in $\Pi$.

Next we show that $d_4$ cannot be used, either, to derive
$\defeasibly(p)$ in $\Pi$. Then $\defeasibly(p)\not\in In(\Pi)=E$,
and we are finished. By the $(-\partial)$ inference condition
there are three cases.

{\em Case 1}: $\forall r\in R_{sd}[p] \exists a\in A(r): -\partial
a \in P(1..i)$. By Induction Hypothesis we conclude that for every
strict or defeasible rule with head $p$ there is at least one
antecedent $a$ such that $\defeasibly(a) \not\in In(\Pi)$.
Therefore, for every possible instantiation of $R$ in default
$d_4$ the prerequisite of $d_4$ is not in $In(\Pi)$. Thus $d_4$
cannot be used to derive $\defeasibly(p)$ in $\Pi$.

{\em Case 2}: $+\Delta\non p \in P(1..i)$. Then no instantiation
of $d_4$ where the consequent is $\defeasibly(p)$ can be applied since
$\non p \in In(\Pi)=E$, by Theorem A.2 (and because $\Pi$ is
successful).

{\em Case 3}: There is $s\in R[\non p]$ such that:

\begin{enumerate}[~~~~~]

\item[(1)] $\forall a\in A(s) +\partial a \in P(1..i)$ and

\item[(2)] $\forall t\in R_{sd}[p]: t\not > s$ or $\exists a\in
A(t): -\partial a \in P(1..i)$

\end{enumerate}
From (1) together with Induction Hypothesis we get:
\[
\drule(s,\non p, [u_1,\ldots,u_n]) \wedge \defeasibly(u_1) \wedge
\ldots\wedge \defeasibly(u_n) \in In(\Pi) \tag{*}
\]
Let $t\in R_{sd}[p]$.

{\em Case 3.1}: $t\not > s$. Then $\mt{sup}(t,s)\not\in W(D)$. So,
$d_6$ with instantiation $t$ for $T$ cannot be used to derive
$\defeated(s,\non p)$ in $\Pi$.

{\em Case 3.2}: $\exists a\in A(t): -\partial a \in P(1..i)$.
Then, by inductive hypothesis, we have $\defeasibly(a) \not\in In(\Pi)$.
Again, $d_6$ with instantiation $t$ for $T$ cannot be used to
derive $\defeated(s,\non p)$ in $\Pi$.

Overall we have shown:
\[
\defeated(s,\non p) \not\in In(\Pi) \tag{**}
\]
From $(*)$ and $(**)$ we get that $d_5$ with instantiation $s$ for
$S$ and $r$ for $R$ can be applied to $In(\Pi)$. Since $\Pi$ is
closed, we conclude $\overruled(r,p)\in In(\Pi)$. Since $r$ was
chosen arbitrarily,  the default $d_4$ cannot be used to prove
$\defeasibly(p)$ in $\Pi$, thus $\defeasibly(p) \not\in E=In(\Pi)$.
\end{proof}

The above two theorems show that if $D$ is decisive, then the stable
model semantics of \M corresponds to the provability in defeasible
logic. However part (c) is not true in the general case, as the
following example shows.

\begin{example}
Consider the defeasible theory

\begin{quote}

$r_1: \ \Rightarrow \neg p$

$r_2: \ p \Rightarrow p$

\end{quote}
In defeasible logic, $+\partial \neg p$ cannot be proven because
we cannot derive $-
\partial p$. However, $\defeasibly(\neg p)$ is a sceptical conclusion of \M
under stable model semantics because it is included in the only
stable model of \M.
\end{example}

\subsection{Embedding under Kunen Semantics}

If we wish to have an equivalence result without the condition of
decisiveness, then we must use a different logic programming
semantics, namely Kunen semantics.

The domain of our interpretation is given by the set of the rule-names
and the set of literals occurring in a defeasible theory $D$.
\begin{itemize}
\item $I(\alpha)=\mathbf{t}$ iff
\begin{enumerate}
\item $\alpha=\mt{fact}(p)$ and $p\in F$;
\item $\alpha=\mt{strict}(r_i,p,[q_1,\dots,q_n])$ and
  $r_i:q_1,\dots,q_n\to p\in R$;
\item $\alpha=\mt{defeasible}(r_i,p,[q_1,\dots,q_n])$ and
  $r_i:q_1,\dots,q_n\Rightarrow p\in R$;
\item $\alpha=\mt{defeater}(r_i,p,[q_1,\dots,q_n])$ and
  $r_i:q_1,\dots,q_n\leadsto p\in R$;
\item $\alpha=\mt{sup}(r_i,r_j)$ and $\langle r_i,r_j\rangle\in>$.
\end{enumerate}

\item $I(\alpha)=\mathbf{f}$ iff
\begin{enumerate}
\item $\alpha=\mt{fact}(p)$ and $p\notin F$;
\item $\alpha=\mt{strict}(r_i,p,[q_1,\dots,q_n])$ and
  $r_i:q_1,\dots,q_n\to p\notin R$;
\item $\alpha=\mt{defeasible}(r_i,p,[q_1,\dots,q_n])$ and
  $r_i:q_1,\dots,q_n\Rightarrow p\notin R$;
\item $\alpha=\mt{defeater}(r_i,p,[q_1,\dots,q_n])$ and
  $r_i:q_1,\dots,q_n\leadsto p\notin R$;
\item $\alpha=\mt{sup}(r_i,r_j)$ and $\langle
r_i,r_j\rangle\notin >$.
\end{enumerate}
\item $I(\alpha)=\mathbf{u}$ otherwise.
\end{itemize}
The intuition behind this interpretation is that the predicates
correspond to the elements of $D$. 

It is immediate to see that 
\[
I(\mt{supportive\_rule}(r_i,p,[\seq{q}]))=
\begin{cases}
   \mathbf{t} & r_i\in R_{sd}[p]\text{ and }\seq{q}=A(r_i)\\
   \mathbf{f} & \text{otherwise}
\end{cases}
\]
Similarly
\[
I(\mt{rule}(r_i,p,[\seq{q}]))=
\begin{cases}
  \mathbf{t} & r\in R[p]\text{ and }\seq{q}=A(r_i)\\
  \textbf{f} & \text{otherwise}
\end{cases}
\]

\begin{theorem} \label{kunen}
\begin{itemize}
  
\item[(a)] $D\vdash +\Delta p \ \Leftrightarrow \ \M\models_K
  \definitely(p)$.
  
\item[(b)]$D\vdash -\Delta p \ \Leftrightarrow \ \M \models_K \neg
  \definitely(p)$.
  
\item[(c)] $D\vdash +\partial p \ \Leftrightarrow \ \M\models_K
  \defeasibly(p)$.
  
\item[(b)]$D\vdash -\partial p \ \Leftrightarrow \ \M \models_K \neg
  \defeasibly(p)$.

\end{itemize}

\end{theorem}

\begin{proof}
\caso{ 1, $\Rightarrow$}
We prove it by induction on the construction of $+\Delta$.

\caso[Inductive base]{} 

$p\in+\Delta^1$ iff $p\in F$ or $\to p\in R$, iff $I(\fact(p))=\Bt$ or
$I(\strict(r,p,[\,]))=\Bt$, so either one of the ground instance
$\definitely(p) \plg \fact(p)$ of the clause~\ref{strictly1} or
$\definitely(p) \plg \strict(r,p,[\,])$ of the clause~\ref{strictly2}
implies $I_1(\definitely(p))=\Bt$.

\caso[Inductive step]{} 
Let us assume that the property holds up
to $n$, and $p\in+\Delta^{n+1}$. This means that either
\begin{enumerate}[~~~~~]
\item $p\in F$, for which we can repeat the same argument as the
  inductive base; or
\item $\exists r\in R_s[p]$, say $r_i$ such that
  $A(r_i)\subseteq+\Delta^{n}$. This implies that, for some $m$
\[
I_m(\strict(r_i,p,[\seq{q}]))=\Bt\ ,
\] 
and, by inductive hypothesis, for each $q_j$  $(1\leq j\leq l)$
\[
I_m(\definitely(q_j))=\Bt\ .
\] 
Thus, by clause~\ref{strictly2},
$I_{m+1}(\definitely(p))=\Bt$.
\end{enumerate}

\caso{ 1, $\Leftarrow$} We use induction on the steps on which
$\definitely(p)$ is supported.

\caso[Inductive base]{}
In this case we have
\[
I_1(\definitely(p))=\Bt \mbox{ iff }
\begin{cases}
  \definitely(p) \plg \fact(p) \mbox{ and } I(\fact(p))=\Bt\\
  \definitely(p) \plg \strict(r,p,[\,]) \mbox{ and }\\
      \phantom{\definitely(p) \plg \fact(p) \mbox{ and}}
      I(\strict(r,p,[\,]))=\Bt
\end{cases}
\]
The first case amounts to $p\in F$, but $F\subseteq\Delta^n$, for any
$n$, therefore $p\in+\Delta_F$. In the second case we have a rule $r$
such that $A(r)=\emptyset$ and so $A(r)\subseteq +\Delta^n$, for any
$n$, hence $p\in+\Delta_F$.

\caso[Inductive step]{} 
Let us assume that the property holds up to $n$, and
\[
I_{n+1}(\definitely(p))=\Bt.
\] 
This is possible if there is either a ground instance of
clause~\ref{strictly1} $\definitely(p)\plg\fact(p)$, for which we
repeat the argument of the inductive base, or a ground instance
\begin{clause}
  $\definitely(p)$:-\\
  \> $\strict(r,p,[\seq{q}])$,\\
  \> $\definitely(q_1)$, \dots, $\definitely(q_n)$.  
\end{clause}
of clause~\ref{strictly2} such that 
\[
I_n(\strict(r,p,[\seq{q}]))=\Bt\ ,
\]
which implies $r\in R_{s}[p]$, and, for each $q_i$,
\[
I_n(\definitely(q_i))=\Bt\ .
\] 
By inductive hypothesis, for some $m,$
$q_i\in+\Delta^m$, so $A(r)\subseteq+\Delta^m$; therefore
$p\in+\Delta^{m+1}$; hence $p\in +\Delta_F$.

\caso{ 2, $\Rightarrow$}
Also in this case we use induction on the construction of $-\Delta$.

\caso[Inductive base]{}
If $p\in-\Delta$ then 
\begin{enumerate}[~~~~~]
\item $p\notin F$, but $p\notin F$ iff $I(\fact(p))=\Bf$;
  this implies that all instances of \ref{strictly1} fails; or
\item $\forall r\in R_s[p](A(r)\cap-\Delta^{0}\neq\emptyset)$. Given
  $-\Delta^0=\emptyset$, the sentence $A(r)\cap-\Delta^{0}\neq\emptyset$
  is always false, therefore the condition can be satisfied only
  vacuously. This means $R_s[p]=\emptyset$, which implies
  \[
  I(\strict(r,p,[\seq{q}]))=\Bf\ ,
  \]
  from which it follows that all ground instances of \ref{strictly2} fail. 
\end{enumerate}
Since $\definitely(p)$ fails in all the cases we conclude
$I_1(\definitely(p))=\Bf$, and
$I_1(\Not~\definitely(p))=\Bt$. 

\caso[Inductive step]{} Let us assume that the property holds up to
$n$, $p\notin-\Delta^n$, and $p\in-\Delta^{n+1}$. This implies that
$p\notin F$, so we can repeat the above argument for the first clause.
If $\forall r\in R_s[p](A(r)\cap\Delta^{n}\neq\emptyset)$ is
satisfied, we have two cases: if it is vacuously we use the argument
of the inductive base, otherwise, for each rule $r$,
$A(r)\cap-\Delta^{n}=q_i$. From this we know that $q_i\in-\Delta^n$,
and we can apply the inductive hypothesis; thus for some $m$,
$I_m(\Not~\definitely(q_i))=\Bt$, then
$I_m(\definitely(q_i))=\Bf$. This happens for each instance
of clause \ref{strictly2}, so $I_{m+1}(\definitely(p))=\Bf$,
therefore $I_{n+1}(\Not~\definitely(p))=\Bt$. 

\caso{ 2, $\Leftarrow$}
The proof is by induction on the step on which $\mt{not~strictly}(p)$ is
supported.

\caso[Inductive base: ]{$n=1$}
\begin{align*}
  I_1(\Not~\definitely(p))=\Bt &\iff
     I_1(\definitely(p))=\Bf\\
     &\iff
     \begin{cases}
       1)\ I(\fact(p))=\Bf,\text{ and}\\
       2)\  \forall q,r, 
       \begin{array}[t]{l}
         I(\strict(r,p,[\seq{q}]),\\
         \phantom{I(xxx}\definitely(q_1),\dots,\\
         \phantom{I(xxx}\definitely(q_l))=\Bf
       \end{array}
     \end{cases}
\end{align*}
1) implies $p\notin F$; for 2), since $\definitely$ is not a primitive
predicate 
\[
I(\definitely(p))=\mathbf{u}
\] 
for every $p$, thus
\[
I(\strict(r,p,[\seq{q}]))=\Bf
\]
for every $q,r$; this implies $R_s[p]=\emptyset$, therefore
$p\in-\Delta^{1}$.

\caso[Inductive step]{} 
Let us assume that it holds up to $n$. We can
repeat the above reasoning, but we have to consider also the case
where for each rule $r_i\in R_{s}[p]$ and some $q_j\in A(r)$,
$I_n(\definitely(q_j))=\Bf$. So $I_n(\Not~\definitely(q_j))=\Bt$; by
inductive hypothesis, $q_j\in-\Delta_F$, then
$A(r)\cap-\Delta_F\neq\emptyset$. Let $-\Delta^m$ be smallest set
satisfying this property. By construction we have $p\in-\Delta^{m+1}$,
hence $p\in-\Delta_F$.

\caso[Cases 3 and 4 ($\Rightarrow$): ]{Inductive base}
The inductive base is trivial since it is easy to verify that both
$+\partial^1$ and $-\partial^1$ are empty.

\caso[Inductive step]{}
We assume that the theorem holds up to $+\partial^n$ and
$-\partial^n$.

\caso{ $p\in +\partial^{n+1}$}
If $p\in +\partial^{n+1}$ because $p\in+\Delta^n$, then we can use the
following ground instance of clause \ref{strictly1}
\begin{clause}
\(
\defeasibly(p)\plg\definitely(p).
\)
\end{clause}
By the inductive hypothesis $I_m(\definitely(p))=\Bt$, for some
appropriate $m$, hence $I_{m+1}(\defeasibly(p))=\Bt$.

Otherwise, there is a rule $r\in R_{sd}[p]$ such that (1)
$A(r)\subseteq +\partial^n$ and $\non p\in-\Delta^n$, and (2) for
every rule $s\in R[\non p]$ either (a) $A(s)\cap
-\partial^n\neq\emptyset$ or (b) there is rule $t\in R_{sd}[p]$ such
that $A(t)\subseteq +\partial^n$ and $t>s$.

Let us consider an appropriate ground instance of clause
\ref{defeasibly2}:
\begin{clause}
  $\defeasibly(p)\plg\suprule(r,p,[\seq{q}])$,\+\\
    $\Not~\defeasibly(\non p)$,\\
    $\defeasibly(q_1),\dots,\defeasibly(q_l),$\\
    $\Not~\overruled(r,p).$
\end{clause}
By construction $I(\suprule(r,p,[\seq{q}]))=\Bt$, and by the inductive
hypothesis and (1), for some appropriate $m$, $I_m(\Not~\definitely(p))=\Bt$
(from $\non p\in-\Delta^n$), $I_m(\defeasibly(q_i))=\Bt$ for all
$q_i\in A(r)$ since $A(r)\subseteq +\partial^n$, Therefore we have to
show that 
\[
I_m(\Not \overruled(r,p))=\Bt\ .
\]
This is equivalent to $I_m(\overruled(r,p))=\Bf$, that means that all
the bodies of the appropriate ground instances of clause
\ref{overruled} are false in $I_m$. By construction for all rules $r\in
R[\non p]$, $I(\drule(s,\non p,[\seq{q^s}]))=\Bt$. Thus we have to
prove that either $I_m(\defeasibly(q^s_i))=\Bf$ or
$I_m(\Not~\defeated(s,\non p))=\Bf$. 

Let us partition $R[\non p]$ in two sets $S'=\set{s\in R[\non p]:
  A(s)\cap -\partial^n=\emptyset}$ and $S''= R[\non p]-S'$. By (2a) and
  the inductive hypothesis we have that for every $s\in S'$
\[
I_m(\defeasibly(q^s_i))=\Bf
\]
for some $q^s_i\in A(s)$.  For the rules in $S''$, on the other hand,
we have to prove that $I_m(\Not~\defeated(s,\non p))=\Bf$.
\[
I_m (\Not~\defeated(s,\non p))=\Bf \mbox{ iff }
I_m (\defeated(s,\non p))=\Bt\ .
\]
Hence we need a ground instance of clause \ref{defeated} evaluated as
true in $I_m$. By (2b) for every rule in $S''$ there is a rule $t\in
R_{sd}[p]$ such that $A(t)\subseteq +\partial^n$ and $t>s$. By
construction we obtain 
\[
I(\suprule(t,p,[\seq{q^t}]))=\Bt \mbox{ and }
I(\suprule(t,s))=\Bt,
\]
and by the inductive hypothesis, for all
$q^t_i\in A(t)$, $I_m(\defeasibly(q^t_i))=\Bt$.

\caso{ $p\in-\partial^{n+1}$}
We have to show that $\Not~\defeasibly(p)$ is supported by the Kunen
semantics. This means that for some $m$,
$I_m(\Not~\defeasibly(p))=\Bt$, which implies
$I_m(\defeasibly(p))=\Bf$. Accordingly we have to show that for each
instance of clauses \ref{defeasibly1} and \ref{defeasibly2} an
element of the body is not supported by the semantics.

Let us examine clause \ref{defeasibly1}:
\begin{clause}
$\defeasibly(p) \plg \definitely(p)$.
\end{clause}
But $-\partial^{n+1}\subseteq-\Delta^{n}$, therefore $p\in-\Delta^n$.
By the inductive hypothesis, for some $m$, $I_m(\Not~\definitely(p))=\Bt$
and $I_m(\definitely(p))=\Bf$.

For clause \ref{defeasibly2}
\begin{clause}
  $\defeasibly(p) \plg$\\
  \> $\suprule(r,p,[\seq{q}]),$\\
  \> $\Not~\definitely(\non p),$\\
  \> $\defeasibly(q_1),\dots,\defeasibly(q_l),$\\
  \> $\Not~\overruled(r,p).$
\end{clause}
Let us examine the cases that imply that $p\in -\partial^{n+1}$.

If $\non p\in+\Delta^n$, then, by the inductive hypothesis
$I_m(\definitely(\non p))=\Bt$ and $I_m(\Not~\definitely(\non p))=\Bf$.

From the basic interpretation it is obvious that we have to consider
only the rules for $p$, since $I(\suprule(r,p,[\seq{q}]))=\Bt$ only if
$r\in R_{sd}[p]$. Let us partition $R_{sd}[p]$ as follows:
$R'=\set{r\in R_{sd}: A(r)\cap -\partial^n\neq\emptyset}$, and
$R''=R-R'$.  

If $r\in R'$, then by the inductive hypothesis, for some $q_i\in
A(r)$, 
\[
I_m(\Not~\defeasibly(q_i))=\Bt\mbox{ and }I_m(\defeasibly(q_{i}))=\Bf.
\]
If $r\in R''$, then we have to show that, for some $m$,
$I_m(\Not~\overruled(r,p))=\Bf$. This means that
$I_m(\overruled(r,p))=\Bt$.

Let us examine again a relevant instance of clause \ref{overruled}:
\begin{clause}
  $\overruled(r,p) \plg$\\
  \> $\drule(s,\non p,[\seq{q^{s}}])$,\\
  \> $\defeasibly(q^s_1),\dots,\defeasibly(q^s_l),$\\
  \> $\Not~\defeated(s,\non p)$.
\end{clause}
Since $p\in-\partial^{n+1}$, there is a rule $s$ such that $s\in
R[\non p]$ and $A(s)\subseteq +\partial^n$. By construction
$I(\drule(s,\non p,[\seq{q^s}]))=\Bt$, and by the inductive hypothesis,
for every $q^{s}_i\in A(s)$:
\[
I_m(\defeasibly(q^s_{i}))=\Bt\ .
\]
Hence we have to show that $I(\Not~\defeated(s,\non p))=\Bt$, i.e.
%which means 
$I(\defeated(s,\non p))=\Bf$, from which we know that all
appropriate substitutions of clause \ref{defeated} fail.
\begin{clause}
  $\defeated(s,\non p)\plg$\\
  \>$\suprule(t,p,[\seq{q^t}])$,\\
  \>$\defeasibly(q^t_1),\dots,\defeasibly(q^t_l)$,\\
  \>$\mt{sup}(t,s)$.
\end{clause}
As we have seen  $R[p]\neq\emptyset$, thus we consider two cases: i)
$t\in R'$, and ii) $t\in R''$.

If $t\in R'$, then, by the inductive hypothesis, there exists a $q^t_i\in
A(t)$ such that $I_m(\defeasibly(q^t_i))=\Bf$.

If $t\in R''$, since $p\in -\partial^{n+1}$, we have that
$t\not>s$, thus $I(\mt{sup}(t,s)=\Bf)$.
We have proved that $I_m(\defeated(s,\non p))=\Bf$, which then imply
$I_m(\overruled(r,p))=\Bt$. Thus, in all cases,
$I_m(\defeasibly(p))=\Bf$ and $I_m(\Not~\defeasibly(p))=\Bt$.  

\caso[Inductive base, ]{$\Leftarrow$}
For the inductive base we consider the literals supported by $I_1$;
however, according to clause \ref{defeasibly1} and clause \ref{defeasibly2},
no literal $\defeasibly(p)$ / $\Not~\defeasibly(p)$ is supported by
$I_1$, thus we consider only $\definitely(p)$ and $\Not~\definitely(p)$.

If $\definitely(p)$ is supported by $I_1$ we have to analyse two cases,
a successful instance of clause \ref{strictly1} 
\begin{clause}
\(
\definitely(p)\plg\fact(p).
\)
\end{clause}
or a successful instance of clause \ref{strictly2} with the following
form:
\begin{clause}
\(
\definitely(p)\plg \strict(r,p,[~])).
\)
\end{clause}
In the first case $I(\fact(p))=\Bt$ and so $p\in F$ and $F\subseteq
+\Delta^m$, for every $m$. In the second case we a rule $r\in R_s[p]$
such that, trivially $A(r)\subseteq +\Delta^0$. Hence $p\in
+\Delta_F$.

On the other hand $I_1(\Not~\definitely(p))=\Bt$ iff
$I_1(\definitely(p))=\Bf$. This means that all appropriate substitution
of the clauses \ref{strictly1} and \ref{strictly2} fail. From clause
\ref{strictly1} we obtain $I(\fact(p))=\Bf$ and so $p\notin F$.  The
only case when \ref{strictly2} fails wrt $I$ is when
$I(\strict(r,p,[\seq{q}]))$ is false for all $r,p,q$. This implies
that $R_s[p]=\emptyset$. It is now immediate to verify that the two
conditions imply that $p\in-\Delta^1$, and by the monotonicity of
$\mathcal{T}$, $p\in -\Delta_F$.

\caso[Inductive step]{}
We assume that the property holds up to $I_n$, and we only show the
cases different from the inductive base.

\caso{ $I_{n+1}(\definitely(p))=\Bt$}
Let us consider the following instance of clause \ref{strictly2}:
\begin{clause}
  $\definitely(p)\plg$\\
  \> $\strict(r,p,[\seq{q}]),$\\
  \> $\definitely(q_1),\dots,\definitely(q_n).$
\end{clause}
Here %we have 
$I_n(\strict(r,p,[\seq{q}]))=\Bt$, and for all $q_i\in
A(r)$, $I_n(\definitely(q_i))=\Bt$.  By construction $r\in R_s[p]$, and
by the inductive hypothesis $A(r)\subseteq +\Delta_F$.

\caso{ $I_{n+1}(\Not~\definitely(p))=\Bf$}
Besides the cases of the inductive base we have to consider all
ground substitutions of 
%clause 
\ref{strictly2} where
$I_n(\strict(r,p,[\seq{q}]))=\Bt$. It follows that  for some $q_i\in
A(r)$, $I_n(\definitely(q_i))=\Bf$. By the inductive hypothesis $q_i\in
-\Delta_F$, hence for some $m$, for all $r\in R_s[p]$, $A(r)\cap
-\Delta^m\neq\emptyset$. Hence $p\in -\Delta_F$.

\caso{ $I_{n+1}(\defeasibly(p))=\Bt$}
We have to consider two cases> The first corresponds to the following
clause:
\begin{clause}
\(
\defeasibly(p)\plg\definitely(p).
\)
\end{clause}
Here $I_n(\definitely(p))=\Bt$, therefore, by the inductive hypothesis
$p\in +\Delta_F$, thus there is an $m$ such that $p\in +\Delta^m$, and
consequently $p\in +\partial^m$, and $p\in +\partial_F$.

Otherwise we have
\begin{clause}
  $\defeasibly(p)\plg \Not~\definitely(\non p),$\\
  \>$\suprule(r,p,[\seq{q^r}]),$\\
  \>$\defeasibly(q^r_1),\dots,\defeasibly(q^r_l),$\\
  \>$\Not~\overruled(r,p).$
\end{clause}
where $I_n(\Not~\definitely(\non p))=\Bt$,
$I_n(\suprule(r,p,[\seq{q^r}]))=\Bt$, and for all $q^r_i$s,
$I_n(\defeasibly(q^r_i))=\Bt$. By construction and the inductive
hypothesis $\non p\in +\Delta_F$, $\exists r\in R_{sd}[p]$, and
$A(r)\subseteq +\partial_F$. Moreover $I_n(\Not~\overruled(r,p))=\Bt$,
and therefore $I_n(\overruled(r,p))=\Bf$. This means that all
substitutions of clause \ref{overruled} are evaluated as false in
$I_n$. 

If $I_n(\drule(s,\non p,[\seq{q^s}]))=\Bf$, then $R[\non
p]=\emptyset$, and then trivially $p\in +\partial_F$.

Otherwise we consider the substitutions such that $I_n(\drule(s,\non
p,[\seq{q^s}]))=\Bt$. For such substitutions either
\[
I_n(\defeasibly(q^s_i))=\Bf,\text{ for some }q^s_i,
\] 
or
\[
I_n(\Not~\defeated(s,\non p))=\Bf.
\]
In the first case, by the inductive hypothesis,
$q^s_i\in-\partial_F$. In the second case $I_n(\defeated(s,\non
q))=\Bt$. This implies $I_n(\suprule(t,p,[\seq{q^t}]))=\Bt$,
$I_n(\defeasibly(q^t_i))=\Bt$ for all $q^t_i\in A(t)$, and
$I_n(\suprule(t>s))=\Bt$. Therefore, by construction and the inductive
hypothesis, we obtain 
\[
\forall s\in R[\non p] \text{ either }A(s)\cap -\partial_F
\neq \emptyset \text{ or } \exists t\in R_{sd}[p]: A(t)\subseteq
+\partial_F, t>s\ .
\]
This implies that there is a suitable $m$ where all those conditions
are satisfied and hence $p\in +\partial^{m+1}$, and consequently
$p\in+\partial_F$.

\caso{ $I_{n+1}(\Not~\defeasibly(p))=\Bt$}
The proof of this case is analogous to the previous case. Indeed the
proof of $\Not~\defeasibly(p)$
corresponds to the constructive negation of $\defeasibly(p)$, and the
conditions for $p\in-\partial^{n+1}$ are the negation  of those for
$p\in +\partial^{n+1}$. 
\end{proof} 

\section{Conclusion}

We motivated and presented a translation of defeasible theories
into logic programs, such that the defeasible conclusions of the
former correspond exactly with the sceptical conclusions of the
latter under the stable model semantics, if a condition of
decisiveness is satisfied. If decisiveness is not satisfied, we
have to use Kunen semantics instead for a complete characterization.

This paper closes an important gap in the theory of nonmonotonic
reasoning, in that it relates defeasible logic with mainstream
semantics of logic programming. This result is particularly
important, since defeasible reasoning is one of the most
successful nonmonotonic reasoning paradigms in applications.

\section*{Acknowledgements}

Preliminary version of the material included in this paper appeared
at AAAI'99 \cite{Maher99} and ICLP'02 \cite{Antoniou02}. 

National ICT Australia is funded by the Australian Government's
Department of Communications, Information Technology and the Arts and
the Australian Research Council through Backing Australia's Ability
and the ICT Centre of Excellence program.

%\bibliographystyle{acmtrans}
%\bibliography{embedding}

\end{document}